\documentclass[journal]{IEEEtran}

\usepackage{cite}
\usepackage{amsmath,amssymb,amsfonts}
\usepackage{algorithmic}
\usepackage{graphicx}
\usepackage{textcomp}
\usepackage{xcolor}
\usepackage{balance}
\usepackage{epsfig}
\usepackage{subfigure}
\usepackage{array}
\usepackage{booktabs}
\usepackage{bbold}
\usepackage{bm}
\usepackage{placeins}
\usepackage{comment}
\usepackage{multirow}
\usepackage{float}
\usepackage{hyperref}
\usepackage{tabularx,colortbl}
\usepackage{mathtools}
\usepackage{enumitem}
\usepackage{algorithm}

\begin{document}
\title{Site-Specific Channel Modeling and Optimization of RIS-Assisted Multiuser MISO Systems
\thanks{This work has been submitted to IEEE for possible publication. Copyright may be transferred without notice, after which this version may no longer be accessible.}}

\author{
Ziqi Liu,~\IEEEmembership{Graduate Student Member,~IEEE},
Wei Yu,~\IEEEmembership{Fellow,~IEEE},
and Sean Victor Hum,~\IEEEmembership{Senior Member,~IEEE}
\thanks{Z. Liu, W. Yu, and S. V. Hum are with the Edward S. Rogers Sr. Department of Electrical and Computer Engineering, University of Toronto, Toronto, ON M5S 3G4, Canada (e-mail: ziqii.liu@mail.utoronto.ca).}
\thanks{This work was supported by the Natural Sciences and Engineering Research Council of Canada (NSERC).}
}
\maketitle
\begin{abstract}
This paper presents a physics-based channel modeling and optimization framework for reconfigurable intelligent surface (RIS)-assisted downlink multi-user multiple-input single-output (MU-MISO) communication systems in site-specific environments. A hybrid ray-tracing (RT) and full-wave electromagnetic analysis approach is developed to construct a deterministic channel model that explicitly captures multipath propagation, RIS scattering behavior, and mutual coupling effects through a non-diagonal load impedance representation. Based on this model, an alternating optimization scheme jointly updates the base-station (BS) beamformer and RIS load impedances to maximize the minimum achievable rate under a total transmit power constraint and practical capacitance limits. The objective of the proposed framework is to provide a reliable initial assessment of the system-level impact of RIS deployment in realistic propagation scenarios. To evaluate this capability, the RIS is operated in a column-paired 1-bit control mode that enables exhaustive evaluation of all realizable configurations in both simulation and measurement. Performance is compared at the distribution level through achievable-rate histograms across all configurations and further examined under small user-location variations. The observed agreement between simulation and measurement demonstrates that the proposed framework reliably captures practical performance trends and provides useful guidance for the design and deployment of RIS-assisted MU-MISO systems in site-specific environments.
\end{abstract}

\begin{IEEEkeywords}
RISs, beyond-5G, MU-MISO, channel modeling, alternating optimization, indoor measurements.
\end{IEEEkeywords}

\section{Introduction}
Multiple-antenna transmission has emerged as a key enabler for achieving high channel capacity and reliability in modern wireless communication networks~\cite{8647620, 8741198, 10379539, BJORNSON20193}. Equipping the base station (BS) with multiple antennas enables wireless systems to exploit spatial multiplexing and simultaneously serve multiple distributed user terminals. Using appropriate beamforming techniques, the BS can direct power toward the intended users while suppressing multi-user interference, thereby exploiting the channel's degrees of freedom (DoF) and improving spectral efficiency~\cite{9246254, mueller2025multi, ali2017beamforming}. Nonetheless, effective precoding relies on accurate channel state information to map each user’s data stream onto a corresponding beamforming vector, where channel information is typically obtained in measurement campaigns through sending pilot signals~\cite{9360709, 10053657}. 
In simulations, statistical channel models such as Rayleigh and Rician fading are commonly used to characterize stochastic channel behavior~\cite{9348156, 8811733}. However, in complex multipath communication environments, such as dense urban or indoor scenarios, transmitting and receiving antennas are often in non-line-of-sight (NLoS) conditions due to obstacle blockages. This can lead to significant propagation losses and shadowed regions. Moreover, the strong multipath scattering between objects makes channel difficult to estimate accurately. While conventional precoding techniques remain effective in mitigating interference, their capability to fully compensate for severe multipath fading and shadowing is limited. To overcome this limitation and improve channel capacities, the concept of smart radio environments has been proposed, aiming to control and enhance wireless communication performance by shaping the propagation paths rather than relying solely on precoding at the transceivers~\cite{renzo2019smart}. In this context, reconfigurable intelligent surfaces (RISs) have emerged as a promising technology that can intelligently reconfigure the wireless environment, thereby improving coverage and capacity in challenging propagation scenarios~\cite{9140329, 10556753}.

RISs are engineered periodic structures composed of sub-wavelength unit cells, each capable of dynamically manipulating electromagnetic waves through controlled reflection, refraction, and/or absorption. These surfaces are typically realized by integrating tunable elements into the unit cells, enabling real-time control of their electromagnetic response. Various tuning mechanisms have been explored, including thermally activated phase-change materials~\cite{wang2016optically,10325447}, mechanically reconfigurable structures~\cite{7934020,10500936}, liquid-crystal materials whose permittivity can be reoriented by electromagnetic fields~\cite{9910023,10732196}, and electronic control using components such as PIN diodes~\cite{7509611,9732917} and varactor diodes~\cite{9551980,9668918}. Among the electronic-control elements, varactors are widely used in microwave applications capable of providing continuous impedance tuning. PIN diodes are more commonly employed at millimeter-wave (mmWave) frequencies, where RIS implementations typically operate in discrete ON/OFF states. With the evolution toward beyond 5G and 6G networks, there is growing interest in operating RISs at mmWave frequencies. Although this work primarily focuses on sub-6~GHz RIS-assisted communication using a varactor-based implementation, the proposed framework can also mimic discrete control by restricting the tunable capacitance to two states, thereby enabling a 1-bit RIS operation for comparison and experimental validation.

In wireless communication networks, deploying RISs in the environment can enhance system performance in two ways. First, as commonly observed in single-input single-output (SISO) scenarios, RISs can improve the received signal strength by intelligently reflecting and focusing incident waves toward the intended users, thereby improving the signal-to-noise ratio (SNR)~\cite{9668918, 9732917}. Second, RISs can introduce additional spatial DoF by unlocking additional independent propagation paths, thereby enabling the parallel transmission of data streams and increasing channel capacity~\cite{9473762, 10623689}. By strategically placing the RIS in the environment and carefully tuning its configuration, the channel matrix can be effectively reshaped to support higher multiplexing gains~\cite{9319694, 9681803}. 
A key challenge in this context lies in the joint optimization of RIS configuration and beamformers to meet objectives, as the total received signal intricately is affected by them both. These two factors are strongly coupled, leading to a nonlinear optimization problem that is difficult to solve using conventional optimization methods. 

To address this challenge, various optimization frameworks have been developed. Heuristic algorithms such as genetic algorithms~\cite{8988246,9295760} and particle swarm optimization~\cite{10639175,11030818} have demonstrated effectiveness in exploring the highly multidimensional search space but often suffer from limited repeatability and may converge to suboptimal local minima. In contrast, deterministic optimization approaches, including convex relaxation-based formulations~\cite{9246254}, block coordinate descent methods~\cite{8982186} and subgradient projection methods~\cite{9681803}, offer improved reproducibility and stable convergence behavior. However, as the number of RIS elements increases, these methods may become computationally intensive. More recently, with the rapid advancement of machine learning, deep learning-based approaches have been proposed to predict the RIS configuration and beamformer directly from deterministic channels obtained from analytical modeling or pilot-based estimation~\cite{9427148,9443516}. Although the above methods demonstrate promising performance gains, most rely on simplified diagonal RIS phase-shift models for the RIS. Such modeling significantly simplifies the optimization process, as the diagonal phase-shift matrix yields an element-wise parametrization of the effective channel in terms of the RIS phase variables. However, in practice, mutual coupling between neighboring RIS elements produces non-negligible electromagnetic interactions, and they can be captured through a non-diagonal coupling model. In this case, the system performance metrics (e.g., achievable rate) become a highly nonlinear function of the tunable load configuration, making the optimization problem considerably more challenging. To address the limitation, \cite{10155675} proposes a non-diagonal formulation intended to capture mutual coupling between RIS elements. It attempts to approximate coupling interactions through additional matrix terms. In reality, however, mutual coupling is characterized by the generalized impedance matrix of the multiport network, which is fixed once the RIS topology, geometry, and material properties are determined~\cite{su2001modeling}. As a result, the coupling among unit cells is an inherent electromagnetic property of the surface and cannot be independently tuned or treated as additional optimization variables.

The novelty of this work lies in extending our previously validated hybrid ray-tracing (RT) and full-wave electromagnetic analysis framework~\cite{liu2025channel} from a single-input scenario to a multi-user, multi-input system. Unlike conventional pilot-based channel estimation approaches, the proposed method explicitly links RIS load impedances to multi-user effective channels through a site-specific physics-based model, thereby enabling joint optimization of the transmit beamformer and RIS configuration. Building on this model, an alternating optimization strategy is developed to iteratively refine both components and maximize the minimum achievable rate under a total BS power constraint. To assess the practical impact of RIS deployment, a hardware-compatible 1-bit RIS architecture is investigated in both simulation and measurement. Rather than requiring exact agreement between simulation- and measurement-optimized configurations, the evaluation focuses on whether the proposed framework can reliably predict system-level performance trends. The results demonstrate that the physics-based modeling and optimization framework consistently captures the relative performance behavior of RIS configurations and provides reliable insight into RIS deployment in realistic site-specific MU-MISO networks.

This paper is organized as follows. Section~\ref{sec:Methodology} presents the proposed hybrid RT and full-wave alternating optimization framework, including the RIS-assisted channel modeling, beamformer design, and RIS load impedance optimization methodology. Section~\ref{sec:Simulation} presents simulation results for a site-specific multi-user multi-antenna communication scenario. Section~\ref{sec:Experiment} reports experimental results and comparisons with simulation, including both configuration-level and statistical performance evaluations to assess the practical fidelity of the proposed framework. Finally, Section~\ref{sec:Conclusion} concludes the paper.

\section{Methodology}
\label{sec:Methodology}
This work considers the downlink of a multi-user multiple-input single-output (MU-MISO) system, where a BS equipped with $M$ transmit antennas serves $K$ single-antenna users through an RIS comprising $N$ tunable unit cells, with $K \leq M$. To model the site-specific propagation environment, the hybrid channel modeling technique, combining the RT and full-wave analysis, is employed to express the total received electric fields as a function of the BS beamforming matrix and the RIS load impedances, forming the basis of the optimization problem. The optimization goal is to maximize the minimum achievable rate across all $K$ users under the total BS power constraint. 

\subsection{Hybrid RT and Full-Wave Framework}
To accurately model the site-specific RIS-assisted channel, a hybrid RT and full-wave analysis framework is developed. RT is used to capture the propagation among the BS antennas, RIS, and user terminals, accounting for spatial geometry, path loss, and multipath effects. Full-wave electromagnetic analysis, through ANSYS HFSS (high-frequency structure simulator), characterizes the RIS scattering matrix under realistic incident fields provided by RT, incorporating mutual coupling, edge diffraction, and truncation effects. Furthermore, the RIS tunable loads are modeled using Thevenin and Norton equivalent circuit~\cite{1203128, 10079112}, where the induced currents on the tunable elements can be expressed in terms of voltages across the load ports and the equivalent impedance representations. Based on the model, the scattered fields of RIS can be decomposed into the two components, which are the fields produced by the unloaded structure and those generated by the tunable elements. In this work, varactor diodes are used as the tunable elements, enabling the use of the circuit-based model. The electric fields received at the user terminals are expressed as
\begin{equation}
    \mathbf{E} = \tilde{\mathbf{H}}\mathbf{W}_{\text{BS}},
    \label{eq:E_channel}
\end{equation}
where $\mathbf{W}_{\text{BS}}\in\mathbb{C}^{M\times K}$ denotes the BS beamforming matrix, which is constrained to satisfy the total transmit power requirement $||\mathbf{W}_{\text{BS}}||^{2}=P_{\text{BS}}$. The matrix $\mathbf{W}_{\text{BS}}$ consists of user-specific beamforming vectors $[\boldsymbol{w}_1,\boldsymbol{w}_2,\ldots,\boldsymbol{w}_K]$, each corresponding to the data stream intended for one user. $\tilde{\mathbf{H}}\in\mathbb{C}^{K\times M}$ is the effective channel matrix, representing the composite channels $[\tilde{\boldsymbol{H}}_1, \tilde{\boldsymbol{H}}_2,\ldots,\tilde{\boldsymbol{H}}_K]^T$ from the BS antennas to all the users given as
\begin{equation}
    \tilde{\mathbf{H}}=\mathbf{H}_u+\mathbf{G}_l^{\text{RT}}\biggl[\operatorname{diag}(\boldsymbol{Z}_{L})- \mathbf{Z}_{ll}\biggr]^{-1}\mathbf{H}_0,
\label{eq:H}
\end{equation}
where $\mathbf{H}_u\in\mathbb{C}^{K\times M}$ represents the baseline BS-user channel including both the direct propagation paths and the scattering contribution of the unloaded RIS, while $\mathbf{H}_0\in\mathbb{C}^{N\times M}$ characterizes the links between the BS transmit antennas and the RIS load ports. The matrix $\mathbf{G}_l^{\text{RT}}\in\mathbb{C}^{K\times N}$ represents the electric-field response from the RIS load ports to the user locations, and $\boldsymbol{Z}_{L}\in\mathbb{C}^{N\times1}$ collects the tunable load impedances. The mutual interactions among RIS elements are captured by the generalized impedance matrix $\mathbf{Z}_{ll}\in\mathbb{C}^{N\times N}$, which accounts for both self- and mutual coupling within the RIS multiport network. All matrices in~\eqref{eq:H} are obtained from the proposed hybrid RT and full-wave framework: $\mathbf{H}_u$, $\mathbf{H}_0$, and $\mathbf{G}_l^{\text{RT}}$ are derived from RT-based field evaluations under appropriate excitations, whereas $\mathbf{Z}_{ll}$ is extracted from a full-wave multiport characterization determined solely by the RIS geometry and material properties.
By substituting~(\ref{eq:H}) into~(\ref{eq:E_channel}), the total received fields can be expressed as a function of the inverse of a non-diagonal impedance matrix coupled with the beamforming matrix, establishing the foundation for optimization. Additionally, although the example system focuses single-antenna users, the proposed hybrid full-wave optimization framework can be extended to multi-antenna Rxs by appropriately redefining $\mathbf{H}_u$, $\mathbf{H}_0$, and $\mathbf{G}_l^{\text{RT}}$.

\subsection{RIS Structure and Operation}
\label{sec:RIS_structure}
\begin{figure}[t!]
    \centering
    \includegraphics[height=0.20\textheight,trim={0cm 1.1cm 15.5cm 1cm},clip]{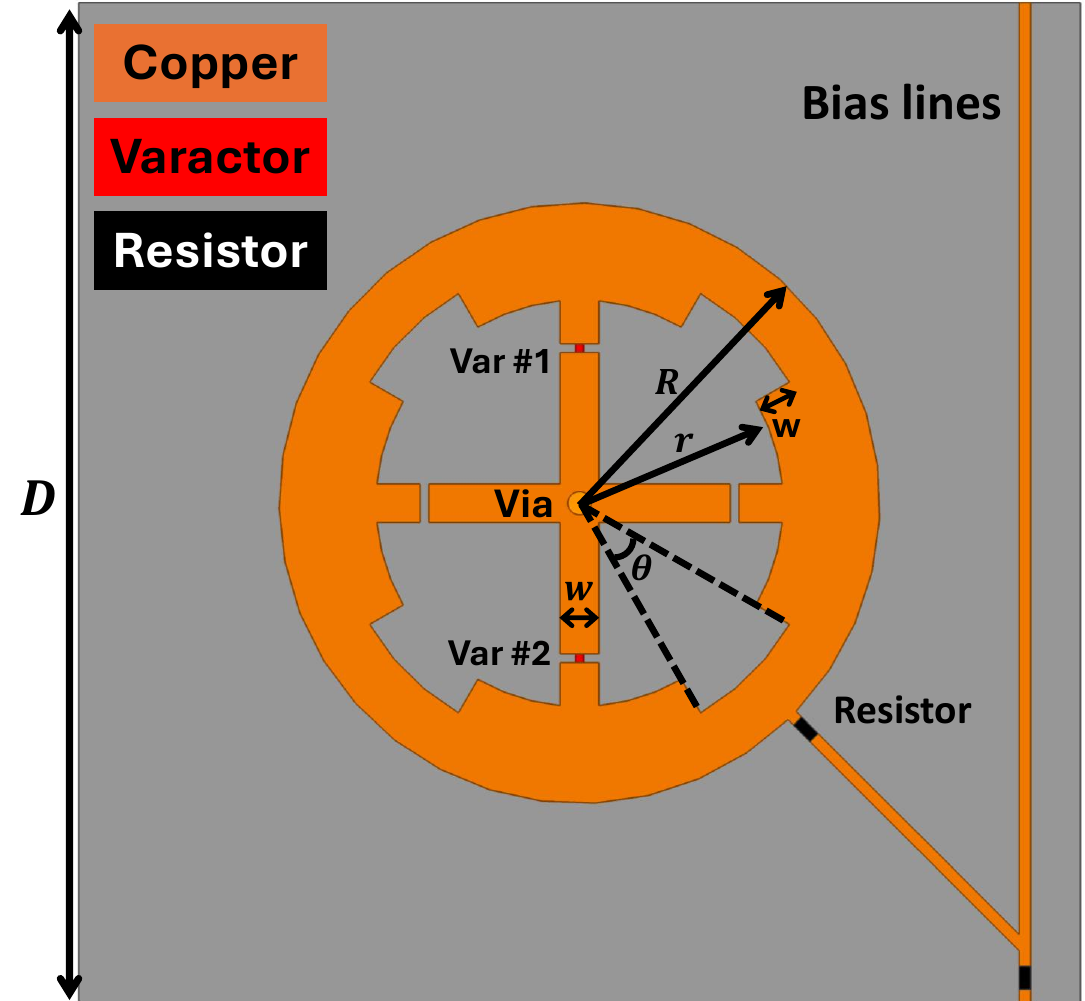}
    \caption{RIS unit cell topology (obtained from~\cite{liu2025channel}). The design features a Jerusalem cross patterned on an RO3003 substrate with two varactors (red) and two resistors (black) enabling reconfigurable operation at 5.8~GHz.}
    \label{fig:Unitcell}
\end{figure}
The RIS structure used in this work is identical to that presented in our previous work~\cite{liu2025channel}. It comprises $20\times11$ unit cells distributed over a total surface area of $10.31\lambda\times6.39\lambda$ ($533.4~\text{mm}\times330.64~\text{mm}$) at 5.8~GHz and is fabricated on a $1.52$~mm-thick RO3003 substrate backed by a ground plane. Each unit cell has a periodicity of $\lambda/2$ ($25.86$~mm). The top copper layer features a patterned structure consisting of a Jerusalem cross, merged with an outer ring, as shown in Fig.~\ref{fig:Unitcell}. A central via connects the metallic patch to the ground plane. Bias lines are soldered along each column of unit cells, and two $10~\text{k}\Omega$ resistors are incorporated in each line to isolate the DC control from the RF signals and to suppress AC leakage. Varactor diodes (MAVR-000120-1411) are mounted on the vertical arms of the Jerusalem cross, enabling reconfigurable control of the transverse electric polarization, where the electric field is aligned with the bias line. Isotropic operation could be achieved by additionally placing varactors on the horizontal arms. However, the dual-polarization control is not considered in this work. Each varactor exhibits an intrinsic resistance $R_v = 2~\Omega$ and inductance $L_v = 0.2$~nH. For the $n$-th RIS element, the capacitance $C_{v,n}$ varies with the applied reverse bias voltage $V_{\text{bias},n}$ according to the standard diode model~\cite{4135599}
\begin{equation}
    C_{v,n}(V_{\text{bias},n}) = \frac{C_J}{\left(1-\frac{V_{\text{bias},n}}{V_J}\right)^m} + C_{\text{par}},
    \label{eq:C_V_equ}
\end{equation}
where $C_J$, $V_J$, $m$, and $C_{\text{par}}$ are experimentally characterized model parameters~\cite{7109870}. Stacking all elements yields the capacitance vector $\boldsymbol{C}_v = [C_{v,1},\dots,C_{v,N}]^T$. The corresponding impedance of the $n$-th tunable load is
\begin{equation}
    Z_{L,n} = R_v + j\omega L_v + \frac{1}{j\omega C_{v,n}},
    \label{eq:Z_loads_element}
\end{equation}
where the load-impedance vector $\boldsymbol{Z}_{L} = [Z_{L,1},\dots,Z_{L,N}]^T \in \mathbb{C}^{N\times1}$ represents the complex impedance values of all RIS unit-cell loads. In this work, the transmit antennas, RIS, and users are positioned on the same plane, such that only two-dimensional wave propagation is considered. Consequently, the RIS is controlled along one dimension, where each column of unit cells shares the same biasing state.

In this work, two RIS operating modes are considered. In the first mode, the RIS employs continuous load capacitance tuning to fully exploit the varactor tuning range and achieve the best simulated performance. In the second mode, the RIS operates in a column-paired 1-bit configuration, where every two adjacent columns share a common state and switch between two discrete responses. The applied bias voltages of $5.02$~V and $3.05$~V correspond to load capacitances of $C_{\mathrm{ON}} = 0.54~\mathrm{pF}$ and $C_{\mathrm{OFF}} = 0.38~\mathrm{pF}$, resulting in approximate unit-cell phase shifts of $-45^\circ$ and $135^\circ$ at $5.8$~GHz. Although this discretized control introduces some performance degradation compared with continuous tuning, it significantly reduces the configuration space and enables exhaustive evaluation of all realizable RIS configurations. Consequently, the achievable-rate performance can be fully characterized in both simulation and measurement, allowing a fair comparison of overall system behavior even when the individually optimal configurations differ, as discussed in Section~\ref{sec:brute-force}.

\subsection{Optimization}
\label{sec:Optimization}
The optimization objective is to determine the BS transmit beamforming matrix $\mathbf{W}_{\text{BS}}$ and the RIS varactor capacitance vector $\boldsymbol{C}_v$ so as to maximize the minimum achievable rate among all users, subject to the total BS power constraint ($P_{\text{BS}}$) and the varactor capacitance tunable range $[C_{\min}, C_{\max}]$. Furthermore, the achievable rate ($R$) serving as the performance metric is defined as
\begin{equation}
    R_k = B\log_2(1+ \text{SINR}_k),
    \label{eq:R_k}
\end{equation}
where $B$ is the bandwidth of the operated system. $\text{SINR}_k$ denotes the signal-to-interference-plus-noise ratio for the $k^\text{th}$ user. The interference represents the power leaked to unintended users, while the noise term, arising from hardware impairments, thermal noise, and other environmental contributions, is modeled as an additive constant. The per-user SINR is computed based on the simulated effective channels as
\begin{equation}
    \text{SINR}_k = \frac{|y_{k,k}(\boldsymbol{w}_k, \boldsymbol{C}_v)|^2}{\sum_{j\neq k} {|y_{k,j}(\boldsymbol{w}_j, \boldsymbol{C}_v)|^2} + \sigma_0^2},
    \label{eq:SINR_k}
\end{equation}
where $y_{k,k} = \tilde{\boldsymbol{H}}_k \boldsymbol{w}_k$ denotes the desired signal received by user $k$. The term $y_{k,j} = \tilde{\boldsymbol{H}}_k \boldsymbol{w}_j$ for $j \in \{1, 2, ..., K\}$ and $j \ne k$ represents the multi-user interference imposed on user $k$ by the beamforming vector intended for user $j$. Here, $\tilde{\boldsymbol{H}}_k$ is the effective channel vector for user $k$ based on all the Tx antennas at the BS, and $\sigma_0^2$ denotes the noise power. 
Since the achievable rate is a monotonically increasing function of SINR, maximizing the achievable rate is equivalent to maximizing the SINR. Therefore, for simplicity, we reformulate the max-min rate optimization problem as a max-min SINR problem, expressed as
\begin{align}
    &\max_{\mathbf{W}_{\text{BS}}, \boldsymbol{C}_v}
    && f(\mathbf{W}_{\text{BS}}, \boldsymbol{C}_v) = \min_{k} \{\text{SINR}_k(\mathbf{W}_{\text{BS}}, \boldsymbol{C}_v)\} \nonumber \\
    &\quad\text{s.t.}
    && \|\mathbf{W}_{\text{BS}}\|^2 = P_{\text{BS}}, \\
    &&& C_{\min} \le C_{v,n} \le C_{\max}, \quad n = 1,\dots, N. \nonumber
\end{align}

The key challenge of this optimization problem lies in the strong nonlinearity of the objective function and the coupling between the optimization variables, denoted $\mathbf{W}_{\text{BS}}$ and $\boldsymbol{C}_v$. To address this, an alternating optimization framework is adopted, where the beamforming matrix and RIS configuration are iteratively refined through an outer-inner loop structure. Specifically, the RIS capacitances are updated in the outer loop using a block coordinate descent (BCD)~\cite{8982186,10364738}. The update direction for each RIS element is computed from the analytical gradient of the minimum SINR ($\text{SINR}_{\min}$) with respect to its capacitance value. In the inner loop, the beamformer $\mathbf{W}_{\text{BS}}$ is optimized using a method based on uplink-downlink duality, ensuring that every evaluation of the objective function employs the optimal transmit strategy for the current RIS state. This steps guarantees a consistent improvement in the minimum SINR across users during the iterative process, while jointly optimizing the RIS configuration and beamformer. It is worth mentioning that a full multidimensional update is not considered since the effective channel depends nonlinearly on the inverse of the non-diagonal impedance matrix as indicated in~(\ref{eq:H}). This mutual coupling causes the gradient components to be strongly interdependent, making a joint update of all load impedances challenging to control. In contrast, the BCD approach updates one block at a time, decomposing the highly nonlinear problem into a sequence of well-conditioned one-dimensional (1D) subproblems. This greatly improves numerical stability and ensures consistent convergence to a stationary point, although global optimality is not guaranteed. The overview of the optimization strategy is summarized in Algorithm~\ref{alg:overview_joint_opt}.
\begin{algorithm}[t!]
\caption{Alternating Optimization of RIS Configuration and BS Beamformer}
\label{alg:overview_joint_opt}
\begin{algorithmic}[1]
\STATE \textbf{Input:} RIS-aided channel components $\mathbf{H}_u$, $\mathbf{H}_0$, $\mathbf{G}_l^{\text{RT}}$, $\mathbf{Z}_{ll}$; BS power budget $P_{\mathrm{BS}}$; convergence tolerance $\epsilon$; maximum iterations $T_{\max}$.
\STATE \textbf{Initialize:} Random RIS capacitance vector $\boldsymbol{C}_v^{(0)} \in [C_{\min},C_{\max}]^N$; set $\mathrm{SINR}_{\min}^{(0)}=-\inf$; set $t\leftarrow 0$.
\STATE Construct RIS-aided channel $\tilde{\mathbf{H}}^{(0)}$ based on  $\boldsymbol{C}_v^{(0)}$, and compute the BS beamformer $\mathbf{W}_{\mathrm{BS}}^{(0)}$ using uplink-downlink duality.
\REPEAT
    \STATE $t \leftarrow t+1$.
    \STATE Set $\boldsymbol{C}_v^{(t)} \leftarrow \boldsymbol{C}_v^{(t-1)}$.
    \FOR{$n=1,\dots,N$}
        \STATE Evaluate gradient $g_n \leftarrow \frac{\partial\,\mathrm{SINR}_{\min}}{\partial C_{v,n}}\big|_{(\boldsymbol{C}_v^{(t)},\,\mathbf{W}_{\mathrm{BS}}^{(t-1)})}$ and choose ascent direction $d_n \leftarrow \mathrm{sign}(g_n)$.
        \STATE Update $C_{v,n}^{(t)} \leftarrow \Pi_{[C_{\min},\,C_{\max}]}\!\Big(C_{v,n}^{(t)} + \alpha_n d_n\Big)$ using Armijo backtracking to select step size $\alpha_n$.
        \STATE Update RIS-aided channel $\tilde{\mathbf{H}}^{(t)}$ using the new $\boldsymbol{C}_v^{(t)}$.
        \STATE Recompute $\mathbf{W}_{\mathrm{BS}}^{(t)}$ via uplink-downlink duality for $\tilde{\mathbf{H}}^{(t)}$ under $P_{\mathrm{BS}}$.
    \ENDFOR
    \STATE Evaluate $\mathrm{SINR}_{\min}^{(t)} \leftarrow \min_k \mathrm{SINR}_k\!\big(\boldsymbol{C}_v^{(t)},\mathbf{W}_{\mathrm{BS}}^{(t)}\big)$.
    \UNTIL{$\left|\mathrm{SINR}_{\min}^{(t)}-\mathrm{SINR}_{\min}^{(t-1)}\right|<\epsilon$ \textbf{or} $t=T_{\max}$}
    \STATE \textbf{Output:} $\boldsymbol{C}_v^\star \leftarrow \boldsymbol{C}_v^{(t)}$, $\mathbf{W}_{\mathrm{BS}}^\star \leftarrow \mathbf{W}_{\mathrm{BS}}^{(t)}$.
\end{algorithmic}
\end{algorithm}

\subsubsection{Beamformer Design Based on Uplink-Downlink Duality}
\label{sec:up_down}
The downlink beamforming optimization under the total BS transmit power constraint is solved using the uplink-downlink duality principle~\cite{730452, 4203115, 4291846, 1453766, 1580783}. In the equivalent virtual uplink, each user $k\in\{1,2,\ldots,K\}$ transmits with power $q_k$, satisfying $\sum_{k=1}^{K} q_k = P_{\text{BS}}$. The effective channel matrix $\tilde{\mathbf{H}}$ is obtained from the hybrid full-wave model, where $\tilde{\boldsymbol{H}}_k$ represents the channel vector from all BS antennas to user $k$. The BS employs a linear minimum mean-square error combiner defined as
\begin{equation}
    \boldsymbol{w}_k^{\text{ul}} = \sqrt{q_k} \left[\sigma_0^2 \mathbf{I} + \sum_{k=1}^{K} q_k \tilde{\boldsymbol{H}}_k^H \tilde{\boldsymbol{H}}_k\right]^{-1}\tilde{\boldsymbol{H}}_k^H,
\end{equation}
where $\boldsymbol{w}_k^{\text{ul}}$ denotes the uplink receive combining vector used at the BS to extract user $k$'s signal across all $M$ antennas, and $(\cdot)^H$ denotes the Hermitian transpose. By stacking the receive combining vectors for all $K$ users, the uplink combining matrix is formed as $\mathbf{W}^{\text{ul}} = \left[\boldsymbol{w}_1^{\text{ul}}, \boldsymbol{w}_2^{\text{ul}},\ldots,\boldsymbol{w}_K^{\text{ul}}\right]\in\mathbb{C}^{M\times K}$.
The uplink SINR for user $k$ is
\begin{equation}
    \text{SINR}_k^{\text{ul}} = \frac{q_k |\tilde{\boldsymbol{H}}_k\boldsymbol{w}_k^{\text{ul}}|^2}{\sum_{j \neq k} q_j |\tilde{\boldsymbol{H}}_j\boldsymbol{w}_k^{\text{ul}}|^2 + \sigma_0^2 \lVert\boldsymbol{w}_k^{\text{ul}}\rVert^2}.
    \label{eq:uplink_SINR}
\end{equation}
To enhance $\text{SINR}_{\min}$ and promote fairness among users, the uplink power allocation is iteratively refined using a fixed-point update inspired by nonlinear Perron–Frobenius theory~\cite{6239604}, expressed as
\begin{equation}
q_k^{\mathrm{new}}
= q_k \frac{\min_i \mathrm{SINR}_i^{\mathrm{ul}}}{\mathrm{SINR}_k^{\mathrm{ul}}}.
\end{equation}
After each update, the power coefficients are normalized to satisfy the uplink sum-power constraint $\sum_k q_k = P_{\mathrm{BS}}$. This iteration reduces the transmit power of users whose SINR exceeds the current minimum, while increasing the power of users with lower SINR, thereby driving the system toward an SINR-balanced operating point. The procedure terminates when the variation in $\mathrm{SINR}_{\min}^{\mathrm{ul}}$ falls below a threshold $\epsilon_u = 10^{-6}$ or when a maximum number of iterations $T_u = 50$ is reached.

After convergence of the virtual uplink optimization, the receive combining matrix $\mathbf{W}^{\text{ul}}$ is obtained. Under the uplink-downlink duality principle~\cite{4489240,7801160}, these combining vectors are used as the initial unnormalized downlink beamformer ($\mathbf{W}^{\text{dl}}$). Furthermore, to equal the total transmit power are identical in the uplink and downlink systems, the downlink must achieve the same SINR as the optimized uplink for each user shown as
\begin{equation}
    \text{SINR}_k^{\text{ul}} = \text{SINR}_k^{\text{dl}},~\forall k.
\end{equation}
To this end, the corresponding downlink power allocation $\boldsymbol{p} = [p_1, p_2, \ldots, p_K]^T$ associated with the beamforming vectors is determined by formulating the downlink per-user SINR expression. By substituting $\text{SINR}^{\text{ul}}$ and $\mathbf{W}^{\text{ul}}$ into the expression, the system of linear equations can be formulated as
\begin{equation}
    p_k G(k,k) - \text{SINR}_k^{\text{ul}}\sum_{j\neq k} p_j G(k,j) = \text{SINR}_k^{\text{ul}}\sigma_0^2,
    \label{eq:SINR_dl}
\end{equation}
where $G(k,j)$ represents the effective downlink channel gain from the beamformer intended for user $j$ to user $k$ shown as
\begin{equation}
    G(k,j) =\left|\tilde{\boldsymbol{H}}_k\boldsymbol{w}_j^{\text{ul}}\right|^2.
    \label{eq:G_kj}
\end{equation}
Diagonal terms $G(k,k)$ represent the desired-signal power, while off-diagonal terms $G(k,j)$ quantify multi-user interference. By substituting (\ref{eq:G_kj}) into (\ref{eq:SINR_dl}), the power allocation ($\boldsymbol{p}$) for each beamforming vector can be calculated. Finally, the downlink beamforming matrix is recovered as
\begin{equation}
    \mathbf{W}^{\text{dl}} = \mathbf{W}^{\text{ul}} \operatorname{diag}\big(\sqrt{\boldsymbol{p}}\big),
\end{equation}
ensuring $\lVert \mathbf{W}^{\text{dl}}\rVert^2 = P_{\text{BS}}$ while preserving SINR equivalence between the uplink and downlink.

\subsubsection{Block Coordinate Descent for RIS Refinement}
\label{sec:BCD}
The RIS configuration is refined using a BCD approach, where each capacitance element $C_{v,n}$ is updated sequentially while all remaining elements and the beamformer remain fixed. For each coordinate, the ascent direction is computed from the analytic derivative of the minimum SINR with respect to $C_{v,n}$, and an Armijo backtracking line search ensures a monotonic improvement of the objective. The gradient of the objective function with respect to $C_{v,n}$ is given by
\begin{equation}
    \bar{g}_n=\frac{\partial\text{SINR}_{k^\star}}{\partial C_{v,n}},
\end{equation}
where the $\text{SINR}_{\min}$ user index is $k^\star = \arg\min_k \text{SINR}_k(\boldsymbol{C}_v)$. The derivative of the received signal component is
\begin{equation}
    \frac{\partial y_{k^\star,j}}{\partial C_{v,n}} = \frac{\partial \tilde{\boldsymbol{H}}_{k^\star}}{\partial C_{v,n}}\boldsymbol{w}_j,
\end{equation}
and the channel derivative follows from the Jacobian of the inverse impedance matrix
\begin{equation}
    \frac{\partial}{\partial C_{v,n}}\mathbf{Z}^{-1} = -\mathbf{Z}^{-1}\left(\frac{\partial \mathbf{Z}}{\partial C_{v,n}}\right)\mathbf{Z}^{-1}.
\end{equation}
Using these relations, the analytic gradient of the minimum SINR is obtained as
\begin{equation}
    \bar{g}_n= \frac{\Delta_{\text{num}}-\text{SINR}_{k^\star}\Delta_{\text{den}}}{\sum_{j\neq k^\star}|y_{k^\star,j}|^2+\sigma_0^2},
\end{equation}
where
\begin{align}
    &\Delta_{\text{num}}=2\text{Re}\left\{y_{k^\star,k^\star}^\ast\frac{\partial y_{k^\star,k^\star}}{\partial C_{v,n}}\right\},\\
    &\Delta_{\text{den}}=2\sum_{j\neq k^\star}\text{Re}\left\{y_{k^\star,j}^\ast\frac{\partial y_{k^\star,j}}{\partial C_{v,n}}\right\}.
\end{align}
Furthermore, to maintain feasibility, gradients that point outside the capacitance bounds $C_{v,n}\in[C_{\min},C_{\max}]$ are suppressed according to
\begin{equation}
\tilde{g}_n=\begin{cases}
    0, & C_{v,n}=C_{\max}~\text{and}~|\bar{g}_n|>0,\\
    0, & C_{v,n}=C_{\min}~\text{and}~|\bar{g}_n|<0,\\
    \bar{g}_n, & \text{otherwise}.
\end{cases}
\end{equation}
The capacitance update is generated as
\begin{equation}
    C_{v,n}^{\text{trial}} = \Pi_{[C_{\min},C_{\max}]}\left(C_{v,n} + \rho_n d_n\right),
\end{equation}
where $d_n={\tilde{g}_n}/{|\tilde{g}_n|}$ is the 1D ascent direction and $\Pi_{[C_{\min},C_{\max}]}$ denotes the Euclidean projection onto the feasible capacitance range. $\rho_0$ is user-selected maximum step size (set to $0.2$~pF in this work). This ensures that the subsequent line search begins with a stable and well-scaled step length.

The trial coordinate $C_{v,n}^{\text{trial}}$ is updated based on the Armijo sufficient-increase condition~\cite[Ch.~3]{nocedal2006numerical} written as
\begin{equation}
    \text{SINR}_{\min}(C_{v,n}^{\text{trial}})\ge\text{SINR}_{\min}(C_{v,n})+\sigma \rho_n \tilde{g}_n d_n,
\end{equation}
where $\sigma\in(0,1)$ is the Armijo parameter. If the condition fails, the step size is reduced as $\rho_n \gets \eta \rho_n$ with $\eta\in(0,1)$, and the trial update is repeated until the step size drops below a prescribed minimum threshold $\rho_{\min}$. The BCD algorithm sequentially updates all capacitances and repeats the sweeps until the improvement in $\text{SINR}_{\min}$ between successive full iterations falls below a threshold $\epsilon_g=10^{-25}$ or a maximum number of iterations $T_g = 50$ is reached. It is worth mentioning that BCD does not guarantee global optimality for this highly nonlinear optimization problem since the BCD method sequentially updates one block of variables at a time based on local gradient information. Nevertheless, it is able to converge to a stationary point for the optimization problem and consistently provides monotonic improvement of the performance.

\subsubsection{Brute-force Configuration Evaluation}
\label{sec:brute-force}
The column-paired 1-bit RIS architecture considered in Section~\ref{sec:RIS_structure} yields a finite and tractable set of configurations, enabling exhaustive evaluation of all possible states. This allows identification of the globally optimal RIS configuration under the 1-bit column-paired constraint and thus provides a benchmark for the ultimate performance achievable in this discrete control mode. Accordingly, a brute-force evaluation approach is adopted in which all realizable RIS configurations are tested. For each candidate configuration, the effective channel is first constructed using the proposed hybrid RT and full-wave framework. The transmit beamformer is then re-optimized via the uplink–downlink duality method described in Section~\ref{sec:up_down}, ensuring that each RIS state is assessed under its best achievable transmission strategy. The resulting minimum achievable rate across users is recorded for every configuration, and the collection of these values forms performance histograms for both simulation and measurement. This distribution-level analysis enables comparison of the overall achievable rate trends induced by RIS deployment without relying on a single optimal configuration. Furthermore, to examine robustness to small spatial perturbations, the same evaluation is performed across multiple nearby user locations, allowing the achievable rate range predicted by the model to be characterized for practical deployment scenarios. The observations can help assess if the proposed modeling and joint optimization framework reliably capture the system-level impact of RIS deployment in realistic site-specific MU-MISO environments.

\section{Synthetic Simulation Examples}
\label{sec:Simulation}
\begin{figure}[t!]
    \centering
    {\includegraphics[width=0.48\textwidth,trim={2.5cm 1.5cm 3.2cm 0.2cm},clip]{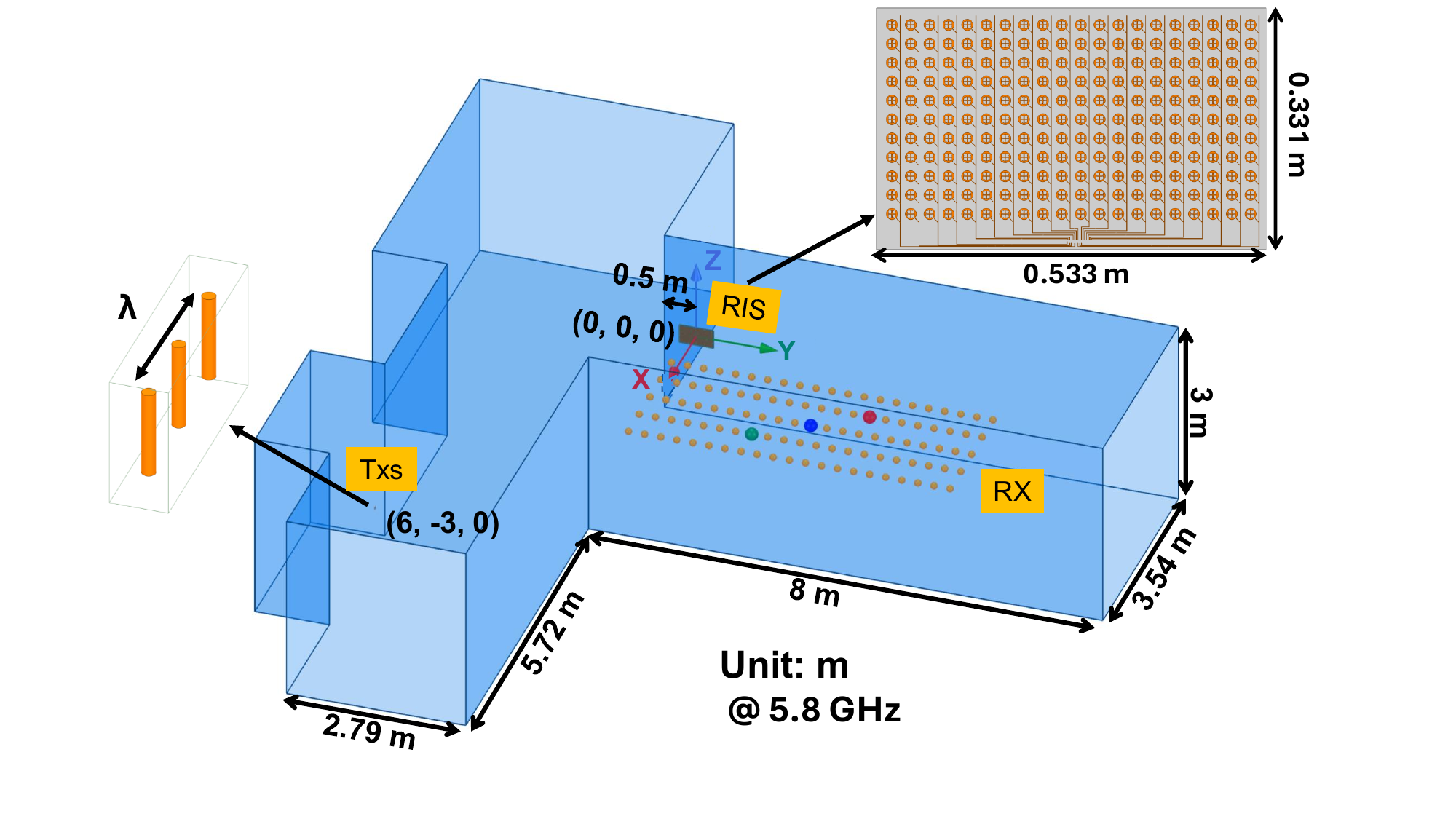}}
    \caption{Illustration of a MU-MISO communication environment with three linear antennas as Txs, each separated by $\lambda/2$ at an operating frequency of $5.8$~GHz. The RIS is located at (0, 0, 0) with 1D control.
    The total electric field distribution is shown within the region indicated by orange dots, and the targeted users are located at (1.3, 3.13, 0)~m, (1.8, 2.38, 0)~m, and (2.3, 1.63, 0)~m, marked as red ,blue, and green dots in order. The RIS structure consists with the one designed in~\cite{liu2025ris, liu2025channel}.
    }\label{fig:SystemSetup}
\end{figure}
We present a MU-MISO example consisting of a single BS and three single-antenna users operating at $5.8$~GHz under NLoS conditions inside a T-shaped corridor on the 8th floor of the Bahen Centre for Information Technology at the University of Toronto, which is the same environment used later for the experimental measurements. The system geometry is illustrated in Fig.~\ref{fig:SystemSetup}. The BS consists three half-wavelength dipole antennas centered at $(6,-3,0)$~m and spaced by $\lambda/2$ at the operating frequency. The RIS is centered at $(0,0,0)$ and has the same topology as the one described in~\cite{liu2025channel, liu2025ris}, having dimensions of $10.30\lambda \times 6.40\lambda$ with $20\times 11$ unit cells. The three user positions are $(1.30,3.13,0)$~m, $(1.80,2.38,0)$~m and $(2.30,1.63,0)$~m, indicated by red, blue and green markers, respectively, in Fig.~\ref{fig:SystemSetup}. In this example, the Tx antennas, RIS and receivers are coplanar. The RIS in controlled in a 1D manner varying along the horizontal direction. Each user is assumed to be served by a single spatial data stream in the MU-MISO transmission model. With three users the BS beamforming matrix is written as
\begin{equation} 
\mathbf{W}_{\text{BS}} = 
\begin{bmatrix} \boldsymbol{w}_{1} ,\boldsymbol{w}_{2} ,\boldsymbol{w}_{3} \end{bmatrix} = 
\begin{bmatrix} {w}_{1,1} & {w}_{2,1} & {w}_{3,1}\\ 
{w}_{1,2} & {w}_{2,2} & {w}_{3,2}\\ 
{w}_{1,3} & {w}_{2,3} & {w}_{3,3} 
\end{bmatrix}, 
\end{equation} 
where $\boldsymbol{w}_{k,m}$ ($k,~m\in\{1,~2,~3\}$) specifies the weight of the $m^\text{th}$ antenna in the beamforming vector $\boldsymbol{w}_{k}$. A total power constraint is imposed at the BS such that $\lVert\mathbf{W}_{\text{BS}}\rVert^2 = P_{\text{BS}}$. 
The noise power is calculated based on the thermal noise,
\begin{equation}
    \sigma_0^2 = kTB,
\end{equation}
where $B$ is the system operating frequency bandwidth, setting to $40$~MHz in this case. $k$ is Boltzmann's constant, and the effective system temperature is taken as $T=900$~K, corresponding to a thermal noise power spectral density of $-169.6$~dBm/Hz.
To investigate the dependence of the minimum achievable rate $R_{\min}$ on the total transmit power, $P_{\text{BS}}$ is varied from $10$~dBm to $30$~dBm in increments of $5$~dBm. Since the measurements are conducted at a single operating frequency, the reported power levels are expressed in dBm/Hz for consistency with measurement results. Additionally, three configurations are examined: (1)~without RIS deployment, (2)~with column-paired RIS under continuous control, and (3)~with RIS under 1-bit control. For the second case, the alternating optimization framework is employed to jointly optimize the beamforming matrix $\mathbf{W}_{\text{BS}}$ and the RIS configuration $\boldsymbol{C}_v$. 

\subsection{No-RIS Case}
In the baseline no-RIS scenario, representing a conventional communication system, the transmit beamformer is optimized based on the deterministic channels obtained from RT. The optimization problem is formulated as
\begin{align}
    \max_{\mathbf{W}_{\text{BS}}} \quad &
        f(\mathbf{W}_{\text{BS}})
        = \min_k \{\mathrm{SINR}_k(\mathbf{W}_{\text{BS}})\}, \nonumber\\
    \text{s.t.} \quad &
        \lVert\mathbf{W}_{\text{BS}}\rVert^2 = P_{\text{BS}}.
\end{align}
The optimized beamformers are obtained using the uplink-downlink duality method described in Section~\ref{sec:up_down}, and the resulting minimum achievable rate $R_{\min}$ is shown by the blue curve in Fig.~\ref{fig:Comparison}. 

For $P_{\text{BS}}=30$~dBm (equivalently $-46.02$~dBm/Hz), the optimized beamformer is
\begin{equation}
    \mathbf{W}_{\text{noRIS}} = \begin{bmatrix}
    0.38e^{-j1.47}&0.54e^{-j0.31}&0.26e^{j0.77}\\
    0.22e^{-j0.08}&0.43e^{j0.85}&0.26e^{j1.47}\\
    0.14e^{j0.30}&0.31e^{j1.16}&0.29e^{j2.11}
    \end{bmatrix}.
    \label{eq:noRIS}
\end{equation}
The resulting minimum achievable rate and average received signal power per Hertz across the three users are $9.13$~bps/Hz and $-141.58$~dBm/Hz, respectively.

\subsection{Continuous-Control Case}
\begin{figure}[t!]
    \centering
    {\includegraphics[width=1\linewidth,trim={0cm 0cm 0cm 0cm},clip]{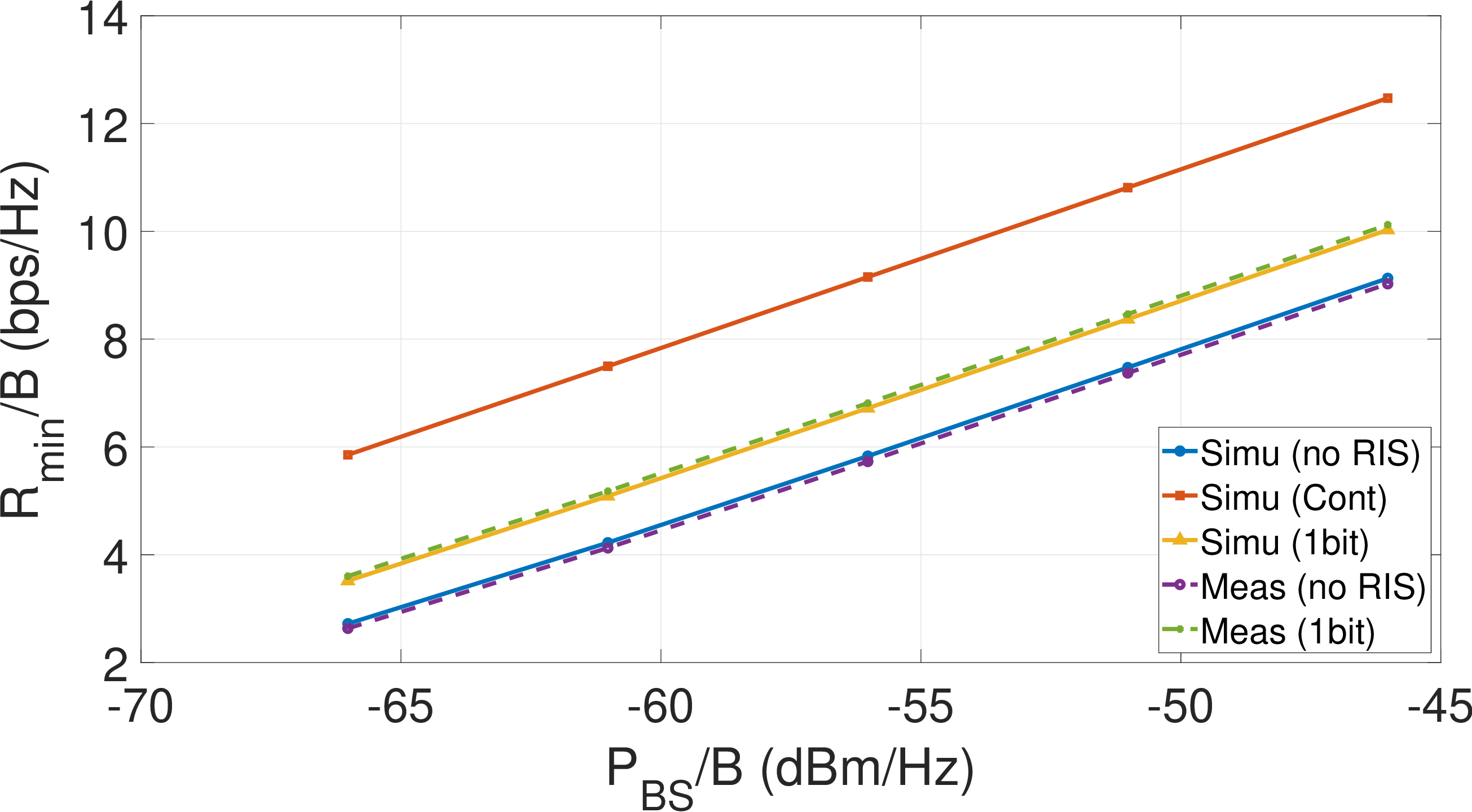}}
    \caption{The comparison between simulation and measurement with and without RIS deployment.}
    \label{fig:Comparison}
\end{figure}

For the continuous-control scenario, each RIS column is tuned independently, and the optimization problem is formulated as
\begin{align}
    \max\limits_{\mathbf{W}_{\text{BS}}, \boldsymbol{C}_v} \quad & 
        f(\mathbf{W}_{\text{BS}}, \boldsymbol{C}_v) = 
        \min_k \{\text{SINR}_k(\mathbf{W}_{\text{BS}}, \boldsymbol{C}_v)\}, \nonumber\\
    \text{s.t.} \quad & \lVert\mathbf{W}_{\text{BS}}\rVert^2 = P_{\text{BS}},\\ 
    & C_{\min} \le C_{v,n} \le C_{\max},\, n\in\{1,2,\ldots,20\}.\nonumber
\end{align}
Based on the alternating optimization framework described in Section~\ref{sec:Optimization}, the RIS configuration and the BS beamformers are refined iteratively in an alternating manner. The RIS update is performed using the BCD scheme, where each $C_{v,n}$ is optimized sequentially. For every coordinate update, an Armijo backtracking line search is imposed to guarantee a sufficient ascent of the objective, with the parameters $\sigma  = 0.05$, $\eta = 0.4$, and $\rho_{\min}=10^{-6}$. After each accepted capacitance update, the BS beamformer is recomputed via uplink-downlink duality, ensuring that every evaluation of the minimum SINR uses the optimal beamforming strategy corresponding to the current RIS state. The values of $R_{\min}$ as a function of $P_{\text{BS}}$ can be seen as the red curve in Fig.~\ref{fig:Comparison}. 

\begin{figure}[t!]
    \centering
    \includegraphics[width=0.48\textwidth,trim={0cm 0cm 0cm 0cm},clip]{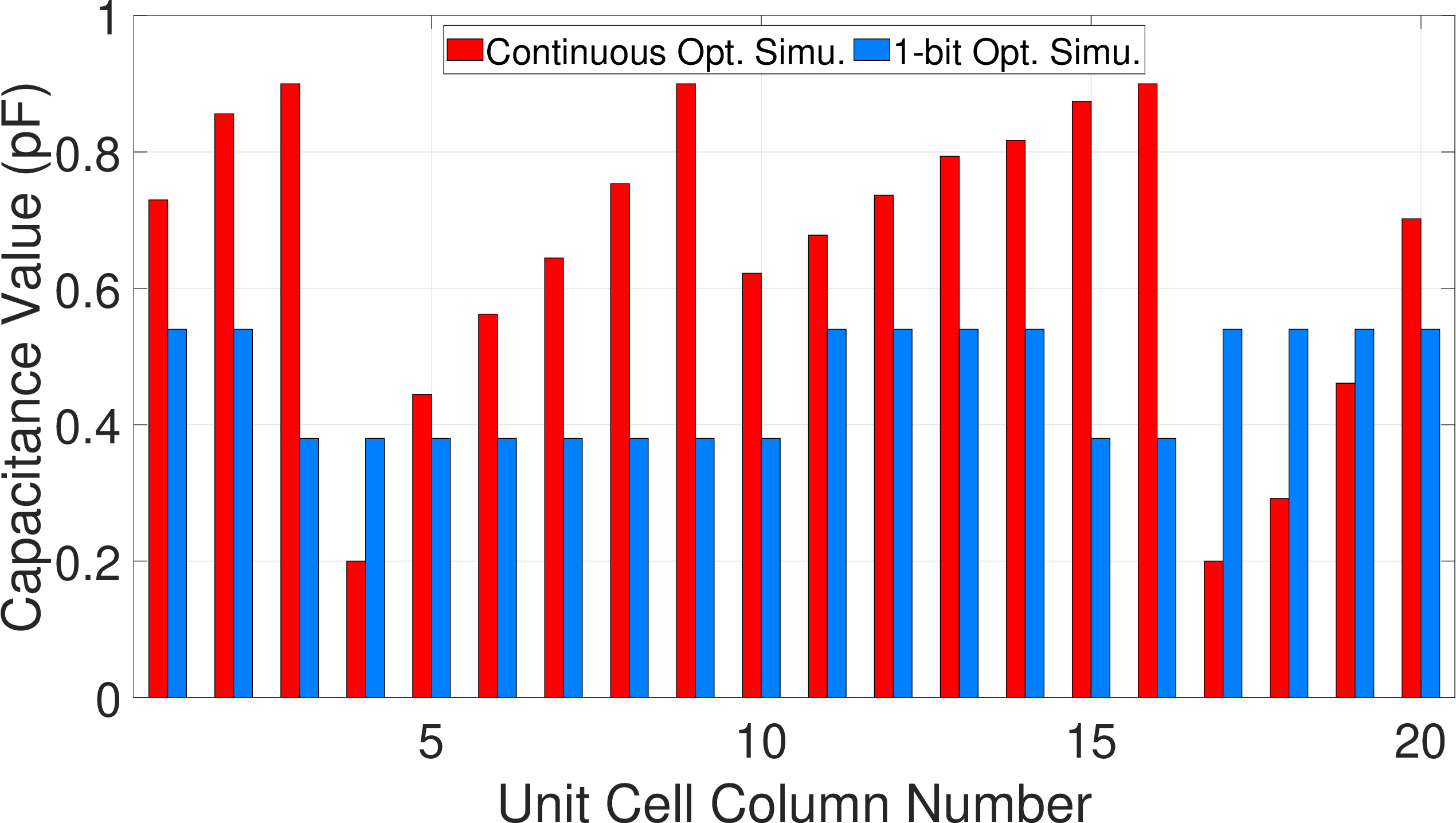}
    \caption{Comparison of RIS capacitance distributions (column-wise) for continuous and 1-bit configurations shown in red and blue bars, respectively.}
    \label{fig:BarPlot_Simu}
\end{figure}

\begin{figure}[t!]
    \centering
    \subfigure[]{\includegraphics[width=0.48\textwidth,trim={0cm 0cm 0cm 0cm},clip]{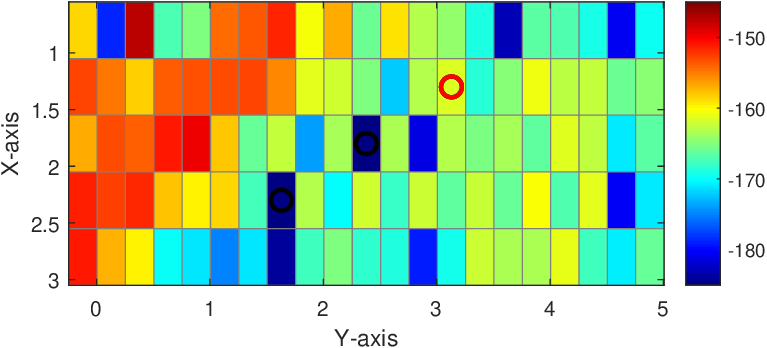}}\\
    \subfigure[]{\includegraphics[width=0.48\textwidth,trim={0cm 0cm 0cm 0cm},clip]{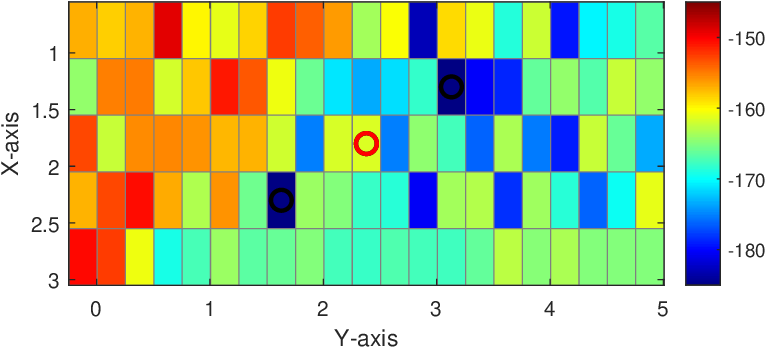}}\\
    \subfigure[]{\includegraphics[width=0.48\textwidth,trim={0cm 0cm 0cm 0cm},clip]{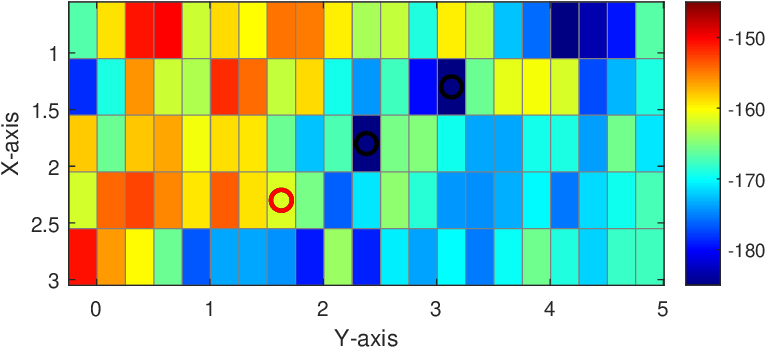}}
    \caption{Signal gain distribution (normalized by $30$~dBm/Hz and shown in dB scale) within the receiving region for three target locations: (1.30, 3.13, 0)~m, (1.80, 2.38, 0)~m, and (2.30, 1.63, 0)~m (red circles). Distributions correspond to $|\mathbf{H}_{\omega_k}|^2$ for $k=1,2,3$ and their corresponding continuously tuned RIS capacitance configurations.}
    \label{fig:MISO_results}
\end{figure}

For $P_{\text{BS}}=30$~dBm/Hz, the optimized beamforming matrix is
\begin{equation}
    \mathbf{W}_{\text{cont}} = \begin{bmatrix}
    0.21e^{j0.83}&0.47e^{j1.56}&0.20e^{-j2.98}\\
    0.42e^{j1.35}&0.50e^{j2.74}&0.22e^{-j2.70}\\
    0.35e^{j3.07}&0.25e^{-j1.04}&0.18e^{j0.60}
    \end{bmatrix}.
\end{equation}
Compared with the no-RIS beamformer, the optimized beamformer with RIS exhibits a more uniform magnitude distribution across antennas, which indicates a more balanced transmission toward different users, leading to more equitable power allocation. The optimized column-wise varactor capacitances are shown by the red bars in Fig.~\ref{fig:BarPlot_Simu}. With the optimized RIS and beamformer, the achievable rate improves to $11.73$~bps/Hz (a gain of $2.6$~bps/Hz relative to the no-RIS case), and the average received power increases to $-133.75$~dBm/Hz (an improvement of $7.83$~dB/Hz over the no-RIS baseline). The signal gain distributions for each beamforming vector with the corresponding the RIS configuration are presented in Fig.~\ref{fig:MISO_results}(a)-(c) for visualization. As shown in the figures, each beamforming vector provides a strong signal at its designated user location while suppressing interference at the other users. Note that the resolution of the field distribution can be edited by adjusting the number of observation points.

\subsection{Column-paired 1-bit RIS}
To provide a physically implementable benchmark for both simulation and measurement, a column-paired 1-bit RIS configuration is implemented. With 10 such column pairs, the RIS yields $2^{10}=1024$ possible binary configurations. For this grouping 1-bit controlled scenario, the optimization variables correspond to the binary capacitance states of these column pairs, and the problem can be expressed as
\begin{align}
    \max_{\mathbf{W}_{\text{BS}}, \boldsymbol{C}_v} \quad &
    f(\mathbf{W}_{\text{BS}}, \boldsymbol{C}_v)
    = \min_{k}\{\text{SINR}_k(\mathbf{W}_{\text{BS}},\boldsymbol{C}_v)\}, \nonumber\\
    \text{s.t.}\quad &
    \|\mathbf{W}_{\text{BS}}\|^2 = P_{\text{BS}},\\
    &
    \boldsymbol{C}_u\in\{C_{\text{ON}},C_{\text{OFF}}\},\quad u=1,\dots,10,\nonumber
\end{align}
where $\boldsymbol{C}_u$ denotes the capacitance state of the $u^{\text{th}}$ column pair. The resulting relationship between the minimum achievable rate $R_{\min}$ and the total transmit power $P_{\text{BS}}$ is shown as the yellow curve in Fig.~\ref{fig:Comparison}, together with the results for the other control modes. As observed, $R_{\min}$ increases approximately linearly with $P_{\text{BS}}$. This behavior arises because the number of users does not exceed the number of transmit antennas ($K \le M$), allowing inter-user interference to be effectively mitigated through zero-forcing precoding~\cite{4153898,9684244}. Consequently, the RIS primarily improves system performance by enhancing the average received signal strength and increasing the effective SNR of the users.

At $P_{\text{BS}}=30$~dBm, the optimized column-wise varactor capacitance values are shown by the blue bars in Fig.~\ref{fig:BarPlot_Simu}. The corresponding optimal beamformer is
\begin{equation}
\mathbf{W}_{\text{1bit}} =
\begin{bmatrix}
0.28e^{-j1.18} & 0.35e^{-j0.13} & 0.16e^{j0.75} \\
0.15e^{j0.07}  & 0.36e^{j0.78}  & 0.29e^{j1.30} \\
0.20e^{j0.01}  & 0.50e^{j1.17}  & 0.49e^{j2.17}
\end{bmatrix}.
\label{eq:1-bit}
\end{equation}
Under this 1-bit constraint, the minimum achievable rate is $10.03$~bps/Hz and the average received power is $-138.87$~dBm/Hz, corresponding to gains of $0.90$~bps/Hz and $2.71$~dB/Hz over the no-RIS baseline, and performance losses of $1.70$~bps/Hz and $5.12$~dB/Hz compared with the continuous-tuning RIS. The minimum achievable rate and average received power for the three cases are summarized in Table~\ref{tab:MISO}. These results indicate that RIS deployment consistently improves both $R_{\min}$ and received signal strength, while the continuous-control mode provides an upper performance bound under the considered environment.

\begin{figure}[t!]
    \centering
    \includegraphics[width=0.98\linewidth,trim={0cm 0cm 0cm 0cm},clip]{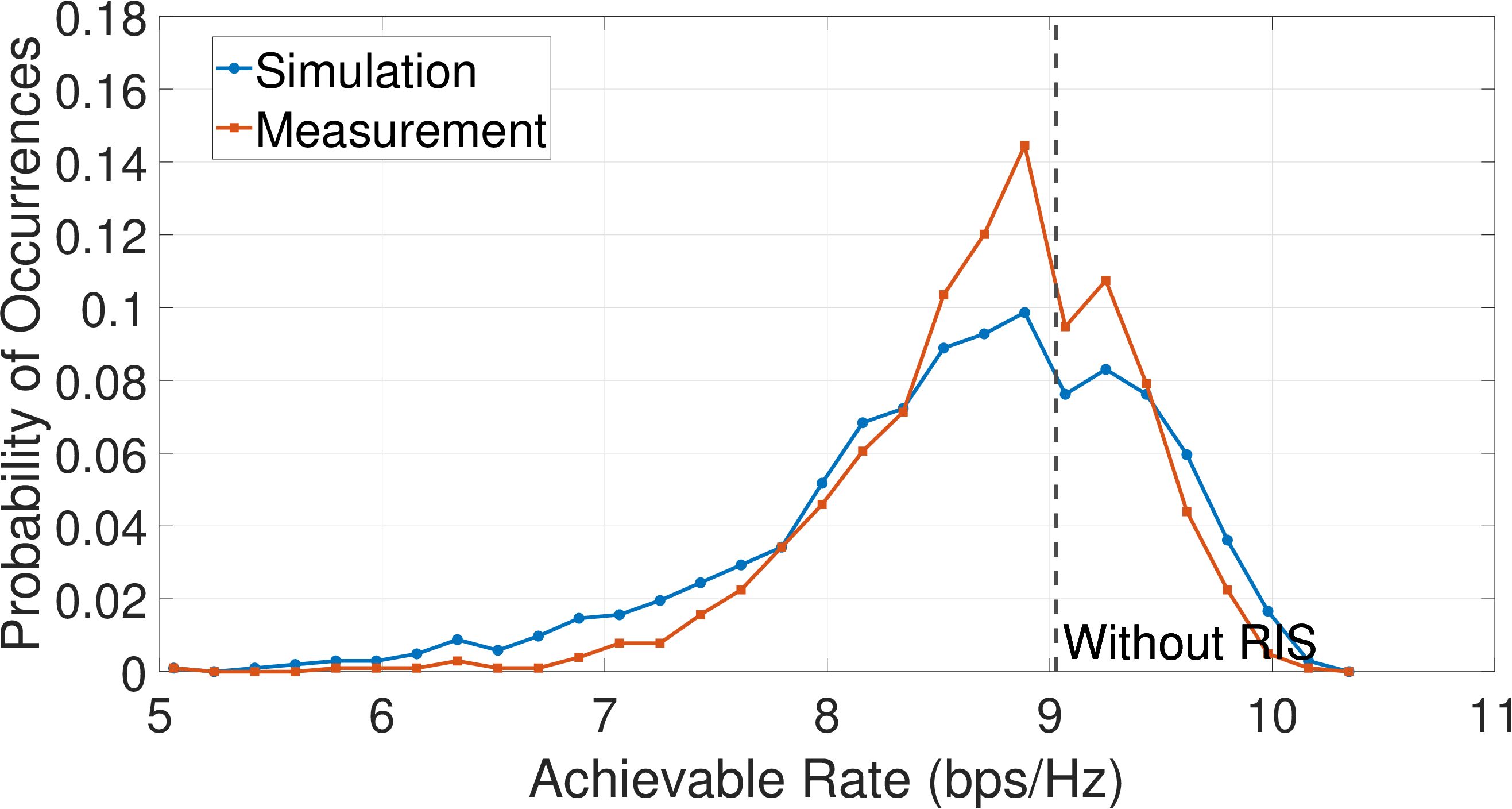}
    \caption{Histogram of minimum achievable rates obtained from exhaustive evaluation of all $2^{10}=1024$ column-paired 1-bit RIS configurations at the designed user locations.}
    \label{fig:Occu_Histogram_Ori}
\end{figure}

\begin{figure}[t!]
    \centering
    \includegraphics[width=0.98\linewidth,trim={0cm 0cm 0cm 0cm},clip]{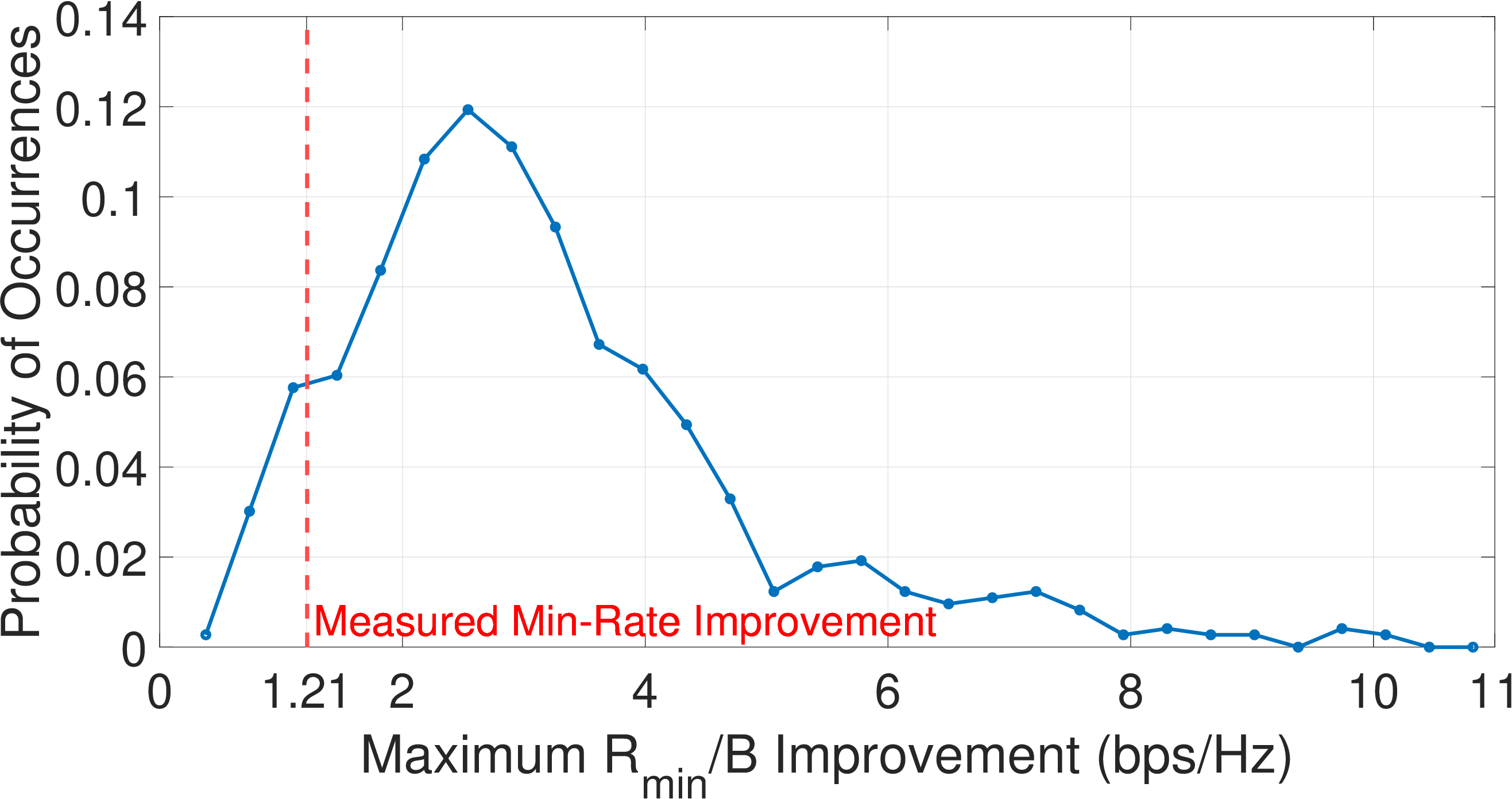}
    \caption{Histogram of minimum achievable-rate improvement achieved by the 1-bit RIS architecture across $729$ nearby user-location combinations, formed by spatial perturbations of $x=\pm0.075$~m and $y=\pm0.092$~m around each designed user location.}
    \label{fig:AchievableRate_Occurrence}
\end{figure}

To further characterize achievable performance under the discrete constraint, the minimum achievable rate is evaluated for all $1024$ configurations, where the transmit beamformer is re-optimized for each configuration. The resulting distribution of minimum achievable rates is illustrated by the blue histogram in Fig.~\ref{fig:Occu_Histogram_Ori}. Approximately $35.06\%$ of the configurations outperform the no-RIS case, indicating that RIS deployment provides performance gains across a broad set of configurations rather than relying on a single optimal state. In addition, the sensitivity of the achievable performance to small user-position variations is investigated to examine the robustness of the RIS-assisted system. For each user, extra 8 nearby sampling points are defined on a rectangular grid centered at the designed user location, spanning offsets of $x=\pm0.075$~m and $y=\pm0.092$~m. One location is selected from each region, resulting in a total of $9\times9\times9=729$ user-location combinations. For each combination, the maximum minimum achievable rate is evaluated, and the resulting rate improvement relative to the no-RIS case is summarized in the histogram shown in Fig.~\ref{fig:AchievableRate_Occurrence}. The results show that the achievable performance improvement remains within a bounded range under small spatial perturbations, indicating that the performance gains provided by the optimized 1-bit RIS can reliably characterize and predict the practical impact of RIS deployment in realistic site-specific MU-MISO networks.

\section{Measurement Results}
\label{sec:Experiment}
\begin{figure*}[t!]
    \centering
    {\includegraphics[width=1\textwidth,trim={0cm 2cm 0cm 3cm},clip]{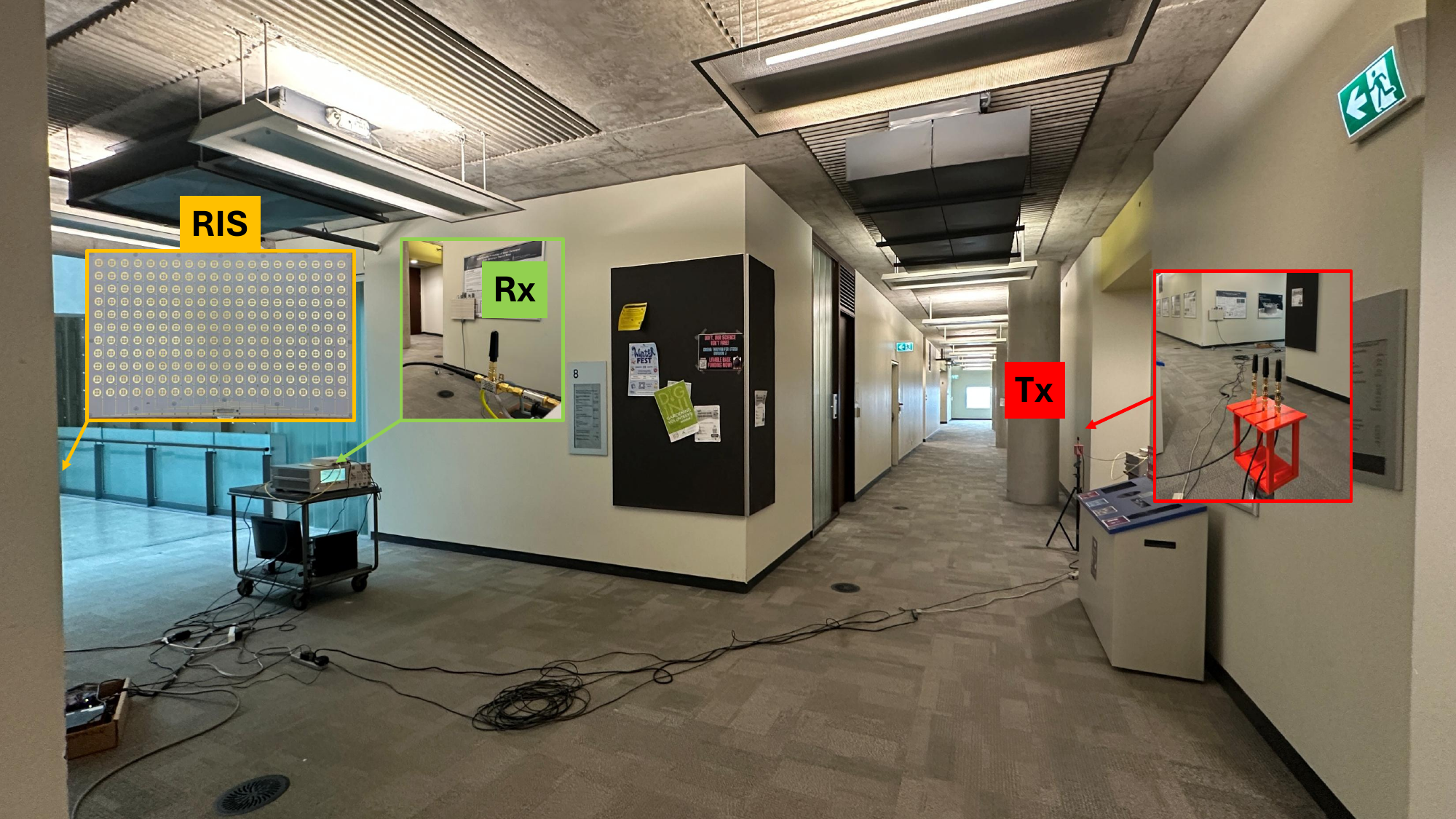}}
    \caption{Measurement system setup in the corridor.}
    \label{fig:Meas}
\end{figure*}
\begin{figure}[t!]
    \centering
    {\includegraphics[width=0.48\textwidth,trim={0.9cm 1.5cm 3.2cm 0.2cm},clip]{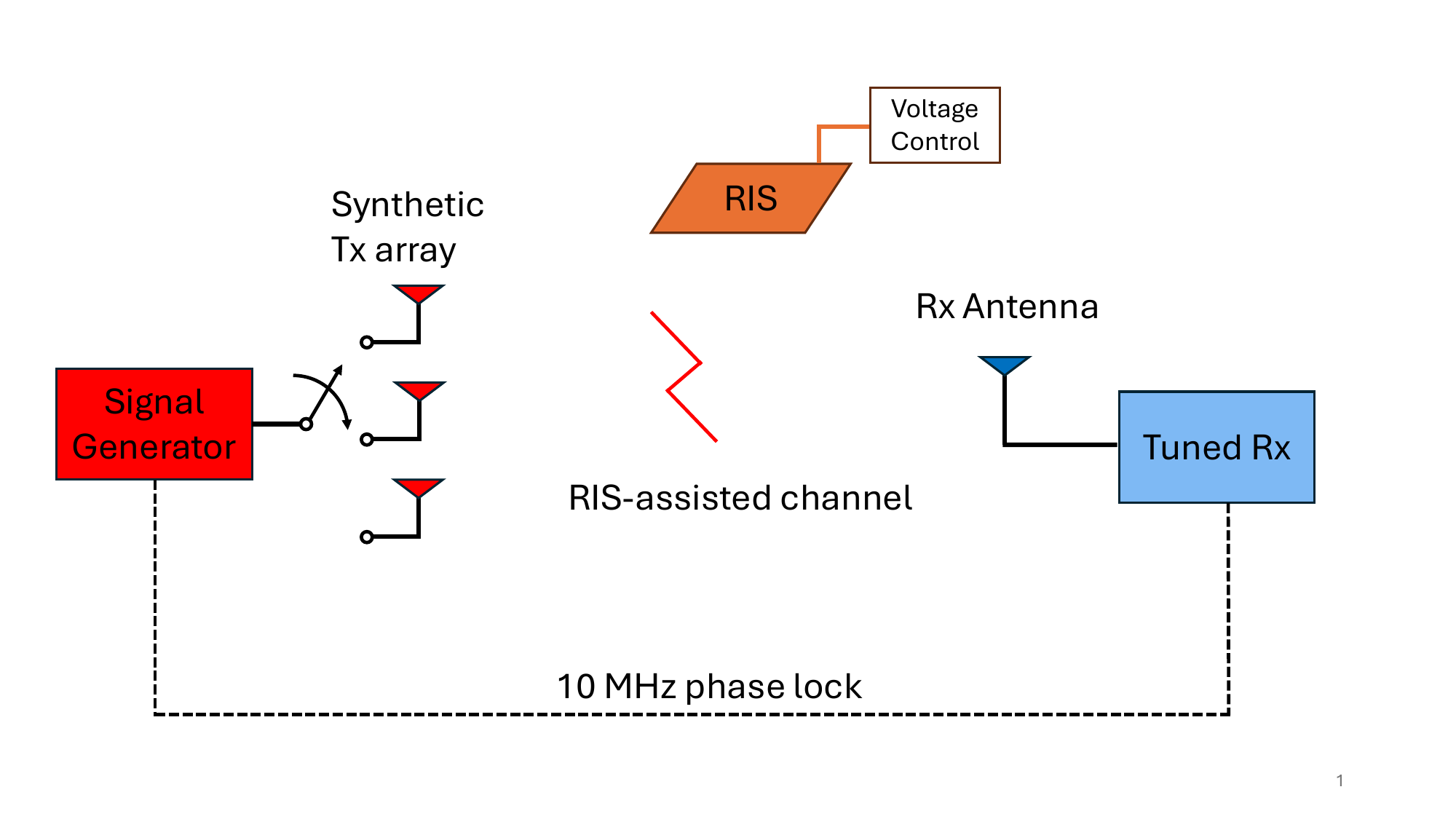}}
    \caption{System block diagram.}
    \label{fig:block_diagram}
\end{figure}
To validate the simulated results, a measurement campaign is conducted in the same corridor environment on the 8th floor of the Bahen Centre for Information Technology at the University of Toronto, as illustrated in Fig.~\ref{fig:Meas}. The physical layout of the measurement environment is consistent with the simulated topology shown in Fig.~\ref{fig:SystemSetup}. In the measurements, the complex channel gain between each element of three-element transmit array and a receiving antenna placed at one of the field points is acquired. The three-antenna synthetic transmit array is controlled by a single-pole multi-throw RF switch. By adjusting the RIS configuration, the scattered fields generated by the RIS can be manipulated and recorded accordingly. The total fields are collected using a vector network analyzer (VNA) functioning as a tuned receiver synchronized to an external signal generator used as a transmitter via a $10$~MHz phase clock to maintain a common phase reference. The receiving antenna is sequentially moved to each user location to capture the corresponding received electric fields, enabling vector (magnitude and phase) measurements of the received signals across the target locations. The overall measurement setup is operated at $5.8$~GHz is depicted in Fig.~\ref{fig:block_diagram}. At this frequency, the phase measurement is highly sensitive ($\lambda \approx 5.17$~cm) such that even sub-wavelength placement errors can produce large phase shifts that alter the performance of both the beamformer and the RIS configuration. Despite this sensitivity, the objective of the proposed hybrid channel modeling and optimization framework is not to predict a single deployable RIS configuration, but rather to provide a reliable initial assessment of the system-level impact of RIS deployment in a site-specific environment. Addtionally, in this measurement, no modulated data signals are transmitted. Instead, a continuous single-tone signal is used to characterize the channel response, allowing accurate acquisition of the complex channel coefficients. By sequentially varying the RIS configurations, the corresponding effective channel matrices are obtained from measurements. The transmit beamformers are then designed during post-processing using the measured channel data, enabling consistent evaluation of achievable-rate performance under different RIS configurations.

Three measurement scenarios are evaluated: (1)~without RIS deployment, (2)~with a continuous-control RIS configuration obtained from simulation, and (3)~with the RIS operating in the column-paired 1-bit mode, where a measurement-driven exhaustive search is applied. For all scenarios, the BS beamformers are redesigned using the measured effective channels. The total transmit power at the BS is effectively swept from $10$~dBm to $30$~dBm in $5$~dBm increments. Since the analysis in this work considers a single-frequency channel representation, the transmit power is expressed in terms of power spectral density by normalizing with respect to the assumed communication bandwidth of $40$~MHz. This corresponds to a transmit power range from $-66.02$~dBm/Hz to $-46.02$~dBm/Hz, with increments of $5$~dBm/Hz. The measured minimum achievable rate $R_{\min}$ for the no-RIS and 1-bit RIS cases is compared with the corresponding simulation results in Fig.~\ref{fig:Comparison}. In this figure, the purple dashed curve represents the measured $R_{\min}$ without the RIS, while the green dashed curve corresponds to the proposed measurement-driven optimization applied to the 1-bit RIS. The close agreement between the measured and simulated curves indicates that the proposed physics-based channel modeling framework provides reliable system-level insight for practical site-specific RIS-assisted MU-MISO systems.

The detailed results at $P_{\text{BS}}/B=-46.02$~dBm/Hz are summarized in Table~\ref{tab:MISO}. When the continuous-control RIS configuration obtained from simulation is directly applied in measurement, only modest improvements are observed, yielding an increase of $0.44$~bps/Hz in $R_{\min}$ and a $1.32$~dB/Hz increase in average received power relative to the no-RIS baseline (from $-141.89$~bps/Hz to $-140.57$~dBm/Hz). These gains are smaller than those predicted in simulation, which can be attributed in part to the richer multipath characteristics of the measured environment that are not fully captured by the simplified RT-based channel model. Additional factors such as fabrication tolerances and placement uncertainties may also contribute to the observed performance gap in practical RIS deployment. For the grouping 1-bit control case, the RIS configuration is determined through exhaustive evaluation of all realizable configurations using the measured effective channel, where the transmit beamformer is re-optimized for each configuration. The histogram of the resulting maximum minimum achievable rates is shown by the red curve in Fig.~\ref{fig:Occu_Histogram_Ori}. The measured distribution closely follows the simulated trend, with $35.35\%$ of the $1024$ configurations outperforming the no-RIS case, compared with $35.06\%$ in simulation. This consistency indicates that, although the configuration yielding the absolute optimum may differ between simulation and measurement, the proposed hybrid framework preserves the relative performance ordering of RIS configurations and accurately predicts the overall achievable performance landscape.

\begin{figure}[t!]
    \centering
    {\includegraphics[width=0.48\textwidth,trim={1.6cm 0cm 0cm 0cm},clip]{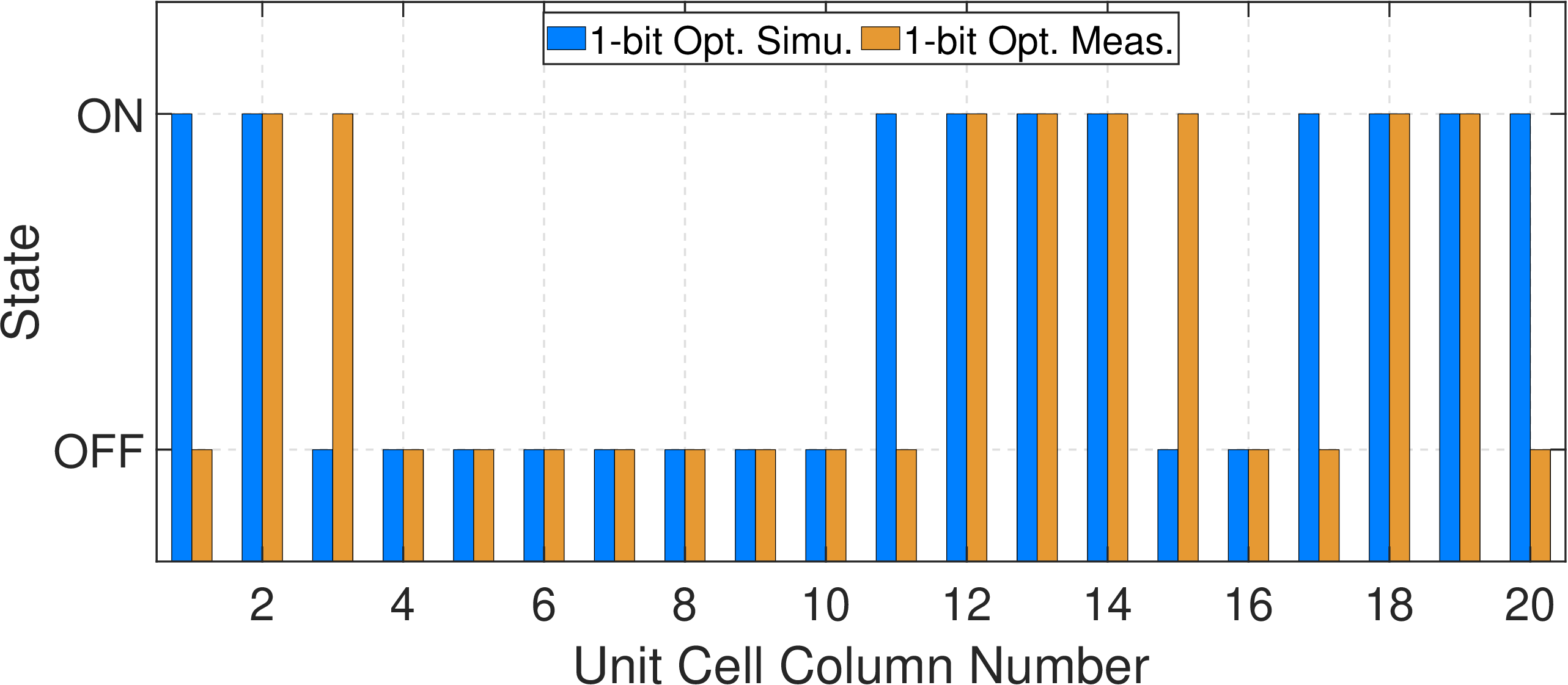}}
    \caption{Comparison of ON/OFF states between simulation and measurement based on brute-force strategy.}
    \label{fig:BarPlot}
\end{figure}

The configuration providing the largest measured rate improvement achieves a gain of $1.11$~bps/Hz over the no-RIS case, as indicated by the red dashed line in Fig.~\ref{fig:AchievableRate_Occurrence}. This improvement lies within the rate improvement range predicted by the spatial perturbation analysis, confirming that the measured optimal performance is consistent with the simulated performance bounds. The corresponding measurement-driven beamformer is
\begin{equation}
    \mathbf{W}_{\text{BS}} = \begin{bmatrix}
    0.40e^{j2.56}&0.30e^{j0.46}&0.51e^{j3.12}\\
    0.49e^{j2.03}&0.30e^{-j1.75}&0.28e^{-j1.01}\\
    0.06e^{j0.71}&0.26e^{-j0.41}&0.10e^{j0.68}
    \end{bmatrix},
\end{equation}
and the RIS load configurations are represented by the orange bars shown in Fig.~\ref{fig:BarPlot}. Although several RIS elements exhibit reversed ON/OFF states compared to the simulation results, the overall switching pattern remains largely consistent. These discrepancies are mainly attributed to systematic phase shifts caused by small placement errors and environmental variations, which alter the incident field distribution at the RIS. Such practical uncertainties, including modeling approximations and fabrication tolerances, are inherent in real deployments. Therefore, the simulation-based optimization framework is intended to provide a physics-grounded prediction of achievable performance trends and configuration-dependent behavior in a given environment. Overall, as summarized in Table~\ref{tab:MISO}, the proposed RIS-assisted framework provides consistent performance improvement over the no-RIS baseline. For the 1-bit configuration, the minimum achievable rate improves by $0.9$~bps/Hz in simulation and $1.11$~bps/Hz in measurement, while the corresponding average received power increases by $2.7$~dB/Hz and $3.3$~dB/Hz, respectively. These results demonstrate that, despite unavoidable discrepancies between simulated and measured models, the proposed hybrid physics-based modeling and optimization framework provides reliable performance prediction and practical guidance for RIS placement and deployment in site-specific MU-MISO systems.
\begin{table}[t!]
\centering
\caption{Representative per-user achievable rate and average received power ($-46.02$~dBm/Hz total power constraint).}
\renewcommand{\arraystretch}{1.2}
\setlength{\tabcolsep}{5pt}
\begin{tabular}{@{} lcc @{}}
\toprule
Configuration & Achievable rate & Avg. received power \\
 & (bps/Hz) & (dBm/Hz)\\
\midrule
\textbf{Simulation} & &\\
No RIS                    & 9.13 & -141.58 \\
Cont. RIS (Optimized) & 11.73 & -133.75 \\
1-bit RIS (Optimized)       & 10.03 & -138.87 \\
\midrule
\textbf{Measurement} & &\\
No RIS                          & 9.03 & -141.89 \\
Cont. RIS (Simulation Config.) & 9.47 & -140.57 \\
1-bit RIS (Measurement-driven)  & 10.12 & -138.59 \\
\bottomrule
\end{tabular}
\label{tab:MISO}
\end{table}

\section{Conclusion}
\label{sec:Conclusion}
This work presented a physics-based channel modeling and joint optimization framework for RIS-assisted MU-MISO communication systems, aiming to maximize the minimum achievable rate through the combined design of the transmit beamforming matrix and RIS configuration. The proposed framework leverages deterministic channel information obtained from a hybrid RT and full-wave electromagnetic analysis at $5.8$~GHz, where the RIS response is represented through a non-diagonal impedance matrix that accounts for mutual coupling among RIS elements, enabling a realistic description of practical surface scattering behavior. In the continuous-control case, the RIS configuration is refined using a BCD approach, while the BS beamformer is updated using uplink-downlink duality under a total transmit power constraint. Simulation results demonstrate that RIS deployment improves both the minimum achievable rate and the received signal strength compared with the no-RIS baseline.

Measurement campaigns conducted in a realistic indoor environment further validate the proposed framework. Due to unavoidable discrepancies between simulated models and practical environments, simulation-optimized RIS configurations do not directly translate to identical optimal configurations in measurement. Instead, exhaustive evaluation of the realizable 1-bit RIS configurations using measured channels enables a physically realizable benchmark for comparison. The results show strong agreement between simulation and measurement at the system level, both in terms of achievable performance trends and in the statistical distribution of achievable rates across RIS configurations. Histogram-based analysis under nearby user-location perturbations further confirms that the measured performance improvements lie within the achievable performance range predicted by simulation, indicating that the proposed framework reliably captures the system-level impact of RIS deployment despite configuration-level mismatches. In the considered MU-MISO setting, where the number of users does not exceed the number of BS antennas, multi-user interference can be effectively mitigated through beamforming, and the RIS primarily contributes by enhancing received signal strength and improving the effective SNR. Owing to its electromagnetic formulation, the proposed framework is inherently frequency scalable and can be extended to other frequency bands and multi-antenna user scenarios. Overall, the presented hybrid modeling and optimization framework provides a reliable tool for evaluating and guiding RIS deployment in realistic site-specific wireless networks.

Future work may investigate more general communication scenarios where the RIS can enhance not only received power but also the effective DoF for the network. Examples include transceiver-pair or multi-user multiple-input multiple-output (MU-MIMO) configurations, and the RIS can facilitate additional spatial multiplexing paths. Furthermore, extending the proposed framework to mmWave frequency bands represents a promising direction for beyond-5G and 6G systems. At these frequencies, varactor-based tuning becomes less practical due to packaging constraints, and alternative tuning technologies such as PIN diodes, $\text{VO}_2$ switches, or liquid crystal materials, typically offering discrete phase states, will be explored. In addition, more advanced optimization techniques can be developed to enhance the robustness of the optimization process and ensure convergence, optimality, and reproducibility. Another important avenue is to generalize the proposed framework for wideband or multi-carrier systems, which will require frequency-dependent modeling and optimization to account for the dispersive behavior of the RIS. Finally, data-driven surrogate models based on neural networks can be introduced to mitigate discrepancies between simulations and measurements, enabling real-time RIS optimization in dynamic environments.

\bibliographystyle{IEEEtran}
\bibliography{IEEEabrv,myref}

\begin{thebibliography}{10}
\providecommand{\url}[1]{#1}
\csname url@samestyle\endcsname
\providecommand{\newblock}{\relax}
\providecommand{\bibinfo}[2]{#2}
\providecommand{\BIBentrySTDinterwordspacing}{\spaceskip=0pt\relax}
\providecommand{\BIBentryALTinterwordstretchfactor}{4}
\providecommand{\BIBentryALTinterwordspacing}{\spaceskip=\fontdimen2\font plus
\BIBentryALTinterwordstretchfactor\fontdimen3\font minus
  \fontdimen4\font\relax}
\providecommand{\BIBforeignlanguage}[2]{{%
\expandafter\ifx\csname l@#1\endcsname\relax
\typeout{** WARNING: IEEEtran.bst: No hyphenation pattern has been}%
\typeout{** loaded for the language `#1'. Using the pattern for}%
\typeout{** the default language instead.}%
\else
\language=\csname l@#1\endcsname
\fi
#2}}
\providecommand{\BIBdecl}{\relax}
\BIBdecl

\bibitem{8647620}
Q.~Wu and R.~Zhang, ``Intelligent reflecting surface enhanced wireless network:
  Joint active and passive beamforming design,'' in \emph{Proc. IEEE Global
  Commun. Conf. (GLOBECOM)}, 2018, pp. 1--6.

\bibitem{8741198}
C.~Huang, A.~Zappone, G.~C. Alexandropoulos, M.~Debbah, and C.~Yuen,
  ``Reconfigurable intelligent surfaces for energy efficiency in wireless
  communication,'' \emph{{IEEE} Trans. Wireless Commun.}, vol.~18, no.~8, pp.
  4157--4170, 2019.

\bibitem{10379539}
Z.~Wang, J.~Zhang, H.~Du, D.~Niyato, S.~Cui, B.~Ai, M.~Debbah, K.~B. Letaief,
  and H.~V. Poor, ``A tutorial on extremely large-scale {MIMO} for 6{G}:
  Fundamentals, signal processing, and applications,'' \emph{IEEE Commun. Surv.
  Tuts.}, vol.~26, no.~3, pp. 1560--1605, 2024.

\bibitem{BJORNSON20193}
E.~Bj\"ornson, L.~Sanguinetti, H.~Wymeersch, J.~Hoydis, and T.~L. Marzetta,
  ``Massive {MIMO} is a reality---what is next?: Five promising research
  directions for antenna arrays,'' \emph{Digit. Signal Process.}, vol.~94, pp.
  3--20, 2019.

\bibitem{9246254}
H.~Xie, J.~Xu, and Y.-F. Liu, ``Max-min fairness in {IRS}-aided multi-cell
  {MISO} systems with joint transmit and reflective beamforming,'' \emph{{IEEE}
  Trans. Wireless Commun.}, vol.~20, no.~2, pp. 1379--1393, 2021.

\bibitem{mueller2025multi}
L.~Mueller, A.~Wolff, S.~Klingel, J.~Krieger, L.~Franke, R.~Stemler, and
  M.~Rahm, ``Multi-user frequency selective beam steering by reconfigurable
  intelligent surfaces in the {K}a-band,'' \emph{Sci. Rep.}, vol.~15, no.~1, p.
  10891, 2025.

\bibitem{ali2017beamforming}
E.~Ali, M.~Ismail, R.~Nordin, and N.~F. Abdulah, ``Beamforming techniques for
  massive {MIMO} systems in 5{G}: overview, classification, and trends for
  future research,'' \emph{Front. Inf. Technol. Electron. Eng.}, vol.~18, pp.
  753--772, 2017.

\bibitem{9360709}
X.~Yuan, Y.-J.~A. Zhang, Y.~Shi, W.~Yan, and H.~Liu,
  ``Reconfigurable-intelligent-surface empowered wireless communications:
  Challenges and opportunities,'' \emph{{IEEE} Wireless Commun.}, vol.~28,
  no.~2, pp. 136--143, 2021.

\bibitem{10053657}
J.~Chen, Y.-C. Liang, H.~V. Cheng, and W.~Yu, ``Channel estimation for
  reconfigurable intelligent surface aided multi-user mm{W}ave {MIMO}
  systems,'' \emph{{IEEE} Trans. Wireless Commun.}, vol.~22, no.~10, pp.
  6853--6869, 2023.

\bibitem{9348156}
T.~Jiang, H.~V. Cheng, and W.~Yu, ``Learning to beamform for intelligent
  reflecting surface with implicit channel estimate,'' in \emph{Proc. IEEE
  Global Commun. Conf. (GLOBECOM)}, 2020, pp. 1--6.

\bibitem{8811733}
Q.~Wu and R.~Zhang, ``Intelligent reflecting surface enhanced wireless network
  via joint active and passive beamforming,'' \emph{{IEEE} Trans. Wireless
  Commun.}, vol.~18, no.~11, pp. 5394--5409, 2019.

\bibitem{renzo2019smart}
M.~Di~Renzo, M.~Debbah, D.-T. Phan-Huy, A.~Zappone, M.-S. Alouini, C.~Yuen,
  V.~Sciancalepore, G.~C. Alexandropoulos, J.~Hoydis, H.~Gacanin \emph{et~al.},
  ``Smart radio environments empowered by reconfigurable {AI} meta-surfaces: An
  idea whose time has come,'' \emph{EURASIP J. Wireless Commun. Netw.}, vol.
  2019, no.~1, pp. 1--20, 2019.

\bibitem{9140329}
M.~Di~Renzo, A.~Zappone, M.~Debbah, M.-S. Alouini, C.~Yuen, J.~de~Rosny, and
  S.~Tretyakov, ``Smart radio environments empowered by reconfigurable
  intelligent surfaces: How it works, state of research, and the road ahead,''
  \emph{{IEEE} J. Sel. Areas Commun.}, vol.~38, no.~11, pp. 2450--2525, 2020.

\bibitem{10556753}
E.~Shi, J.~Zhang, H.~Du, B.~Ai, C.~Yuen, D.~Niyato, K.~B. Letaief, and X.~Shen,
  ``{RIS}-aided cell-free massive {MIMO} systems for 6{G}: Fundamentals, system
  design, and applications,'' \emph{Proc. {IEEE}}, vol. 112, no.~4, pp.
  331--364, 2024.

\bibitem{wang2016optically}
Q.~Wang, E.~T.~F. Rogers, B.~Gholipour, C.-M. Wang, G.~Yuan, J.~Teng, and N.~I.
  Zheludev, ``Optically reconfigurable metasurfaces and photonic devices based
  on phase change materials,'' \emph{Nat. Photon.}, vol.~10, no.~1, pp. 60--65,
  2016.

\bibitem{10325447}
H.~Wang, Z.~Zhang, B.~Zhu, J.~Dang, and L.~Wu, ``Optical reconfigurable
  intelligent surfaces aided optical wireless communications: Opportunities,
  challenges, and trends,'' \emph{{IEEE} Wireless Commun.}, vol.~30, no.~5, pp.
  28--35, 2023.

\bibitem{7934020}
X.~Yang, S.~Xu, F.~Yang, M.~Li, Y.~Hou, S.~Jiang, and L.~Liu, ``A broadband
  high-efficiency reconfigurable reflectarray antenna using mechanically
  rotational elements,'' \emph{{IEEE} Trans. Antennas Propag.}, vol.~65, no.~8,
  pp. 3959--3966, 2017.

\bibitem{10500936}
M.~Baena-Molina, {\'A}.~Palomares-Caballero, G.~Martinez-Garcia,
  R.~Padial-Allu{\'e}, P.~Padilla, and J.~F. Valenzuela-Vald{\'e}s, ``1-bit
  {RIS} unit cell with mechanical reconfiguration at 28 {GHz},'' in \emph{Proc.
  Eur. Conf. Antennas Propag. (EuCAP)}, 2024, pp. 1--5.

\bibitem{9910023}
S.~Aboagye, A.~R. Ndjiongue, T.~M.~N. Ngatched, and O.~A. Dobre, ``Design and
  optimization of liquid crystal {RIS}-based visible light communication
  receivers,'' \emph{{IEEE} Photon. J.}, vol.~14, no.~6, pp. 1--7, 2022.

\bibitem{10732196}
A.~Hojjati-Firoozabadi and R.~Mansour, ``A novel {RIS} unit cell design
  enabling seamless integration with {VO}$_2$ switches,'' in \emph{Proc. Eur.
  Microw. Conf. (EuMC)}, 2024, pp. 896--899.

\bibitem{7509611}
H.~Yang, F.~Yang, S.~Xu, M.~Li, X.~Cao, and J.~Gao, ``A 1-bit multipolarization
  reflectarray element for reconfigurable large-aperture antennas,''
  \emph{{IEEE} Antennas Wireless Propag. Lett.}, vol.~16, pp. 581--584, 2017.

\bibitem{9732917}
G.~C. Trichopoulos, P.~Theofanopoulos, B.~Kashyap, A.~Shekhawat, A.~Modi,
  T.~Osman, S.~Kumar, A.~Sengar, A.~Chang, and A.~Alkhateeb, ``Design and
  evaluation of reconfigurable intelligent surfaces in real-world
  environment,'' \emph{IEEE Open J. Commun. Soc.}, vol.~3, pp. 462--474, 2022.

\bibitem{9551980}
X.~Pei, H.~Yin, L.~Tan, L.~Cao, Z.~Li, K.~Wang, K.~Zhang, and E.~Björnson,
  ``{RIS}-aided wireless communications: Prototyping, adaptive beamforming, and
  indoor/outdoor field trials,'' \emph{{IEEE} Trans. Commun.}, vol.~69, no.~12,
  pp. 8627--8640, 2021.

\bibitem{9668918}
A.~Araghi, M.~Khalily, M.~Safaei, A.~Bagheri, V.~Singh, F.~Wang, and
  R.~Tafazolli, ``Reconfigurable intelligent surface ({RIS}) in the sub-6 {GH}z
  band: {D}esign, implementation, and real-world demonstration,'' \emph{{IEEE}
  Access}, vol.~10, pp. 2646--2655, 2022.

\bibitem{9473762}
M.~Fu, Y.~Zhou, and Y.~Shi, ``Reconfigurable intelligent surface for
  interference alignment in {MIMO} device-to-device networks,'' in \emph{Proc.
  IEEE Int. Conf. Commun. Workshops (ICC Workshops)}, 2021, pp. 1--6.

\bibitem{10623689}
{\"O}.~T. Demir and E.~Bj{\"o}rnson, ``Wideband channel capacity maximization
  with beyond diagonal {RIS} reflection matrices,'' \emph{{IEEE} Wireless
  Commun. Lett.}, vol.~13, no.~10, pp. 2687--2691, 2024.

\bibitem{9319694}
G.~Gradoni and M.~Di~Renzo, ``End-to-end mutual coupling aware communication
  model for reconfigurable intelligent surfaces: An electromagnetic-compliant
  approach based on mutual impedances,'' \emph{{IEEE} Wireless Commun. Lett.},
  vol.~10, no.~5, pp. 938--942, 2021.

\bibitem{9681803}
T.~Jiang and W.~Yu, ``Interference nulling using reconfigurable intelligent
  surface,'' \emph{{IEEE} J. Sel. Areas Commun.}, vol.~40, no.~5, pp.
  1392--1406, 2022.

\bibitem{8988246}
T.~Shan, X.~Pan, M.~Li, S.~Xu, and F.~Yang, ``Coding programmable metasurfaces
  based on deep learning techniques,'' \emph{{IEEE} Trans. Emerg. Sel. Topics
  Circuits Syst.}, vol.~10, no.~1, pp. 114--125, 2020.

\bibitem{9295760}
K.~Zhang, C.~Liu, H.~Wang, and Y.~Song, ``An {IRS}-aided mmwave massive {MIMO}
  systems based on genetic algorithm,'' in \emph{Proc. Int. Conf. Commun.
  Technol. (ICCT)}, 2020, pp. 288--293.

\bibitem{10639175}
V.~D. Pegorara~Souto, R.~Demo~Souza, and B.~F. Uch\^oa-Filho, ``{PSO}-based
  optimization of {STAR-RIS} aided {NOMA} wireless communication networks,'' in
  \emph{Proc. Int. Symp. Wireless Commun. Syst. (ISWCS)}, 2024, pp. 1--6.

\bibitem{11030818}
A.~Irshad, A.~Kosasih, V.~Petrov, and E.~Björnson, ``Pre-optimized irregular
  arrays versus movable antennas in multi-user {MIMO} systems,'' \emph{{IEEE}
  Wireless Commun. Lett.}, vol.~14, no.~8, pp. 2656--2660, 2025.

\bibitem{8982186}
H.~Guo, Y.-C. Liang, J.~Chen, and E.~G. Larsson, ``Weighted sum-rate
  maximization for reconfigurable intelligent surface aided wireless
  networks,'' \emph{{IEEE} Trans. Wireless Commun.}, vol.~19, no.~5, pp.
  3064--3076, 2020.

\bibitem{9427148}
T.~Jiang, H.~V. Cheng, and W.~Yu, ``Learning to reflect and to beamform for
  intelligent reflecting surface with implicit channel estimation,''
  \emph{{IEEE} J. Sel. Areas Commun.}, vol.~39, no.~7, pp. 1931--1945, 2021.

\bibitem{9443516}
O.~\"Ozdogan and E.~Bj\"ornson, ``Deep learning-based phase reconfiguration for
  intelligent reflecting surfaces,'' in \emph{Proc. Asilomar Conf. Signals,
  Syst., Comput.}, 2020, pp. 707--711.

\bibitem{10155675}
M.~Nerini, S.~Shen, and B.~Clerckx, ``Closed-form global optimization of beyond
  diagonal reconfigurable intelligent surfaces,'' \emph{{IEEE} Trans. Wireless
  Commun.}, vol.~23, no.~2, pp. 1037--1051, 2024.

\bibitem{su2001modeling}
T.~Su and H.~Ling, ``On modeling mutual coupling in antenna arrays using the
  coupling matrix,'' \emph{Microw. Opt. Technol. Lett.}, vol.~28, no.~4, pp.
  231--237, 2001.

\bibitem{liu2025channel}
Z.~Liu and S.~V. Hum, ``Channel-informed {RIS} analysis and optimization using
  hybrid ray-tracing and full-wave simulation framework,'' \emph{{IEEE} Trans.
  Antennas Propag.}, vol.~74, no.~1, pp. 909--921, 2026.

\bibitem{1203128}
R.~Collin, ``Limitations of the {T}hevenin and {N}orton equivalent circuits for
  a receiving antenna,'' \emph{{IEEE} Antennas Propag. Mag.}, vol.~45, no.~2,
  pp. 119--124, 2003.

\bibitem{10079112}
S.~K.~R. Vuyyuru, R.~Valkonen, D.-H. Kwon, and S.~A. Tretyakov, ``Efficient
  anomalous reflector design using array antenna scattering synthesis,''
  \emph{{IEEE} Antennas Wireless Propag. Lett.}, vol.~22, no.~7, pp.
  1711--1715, 2023.

\bibitem{4135599}
V.~I. Cojocaru and T.~J. Brazil, ``A large-signal equivalent circuit model for
  hyperabrupt p-n junction varactor diodes,'' in \emph{Proc. Eur. Microw.
  Conf.}, 1992, pp. 1115--1121.

\bibitem{7109870}
N.~Nguyen-Trong, T.~Kaufmann, L.~Hall, and C.~Fumeaux, ``Analysis and design of
  a reconfigurable antenna based on half-mode substrate-integrated cavity,''
  \emph{{IEEE} Trans. Antennas Propag.}, vol.~63, no.~8, pp. 3345--3353, 2015.

\bibitem{10364738}
Y.~Zhou, Y.~Liu, H.~Li, Q.~Wu, S.~Shen, and B.~Clerckx, ``Optimizing power
  consumption, energy efficiency, and sum-rate using beyond diagonal {RIS}—a
  unified approach,'' \emph{{IEEE} Trans. Wireless Commun.}, vol.~23, no.~7,
  pp. 7423--7438, 2024.

\bibitem{730452}
F.~Rashid-Farrokhi, K.~R. Liu, and L.~Tassiulas, ``Transmit beamforming and
  power control for cellular wireless systems,'' \emph{{IEEE} J. Sel. Areas
  Commun.}, vol.~16, no.~8, pp. 1437--1450, 1998.

\bibitem{4203115}
W.~Yu and T.~Lan, ``Transmitter optimization for the multi-antenna downlink
  with per-antenna power constraints,'' \emph{{IEEE} Trans. Signal Process.},
  vol.~55, no.~6, pp. 2646--2660, 2007.

\bibitem{4291846}
M.~Codreanu, A.~Tolli, M.~Juntti, and M.~Latva-aho, ``Joint design of {T}x-{R}x
  beamformers in {MIMO} downlink channel,'' \emph{{IEEE} Trans. Signal
  Process.}, vol.~55, no.~9, pp. 4639--4655, 2007.

\bibitem{1453766}
M.~Schubert and H.~Boche, ``Iterative multiuser uplink and downlink beamforming
  under {SINR} constraints,'' \emph{{IEEE} Trans. Signal Process.}, vol.~53,
  no.~7, pp. 2324--2334, 2005.

\bibitem{1580783}
W.~Yu, ``Uplink-downlink duality via minimax duality,'' \emph{{IEEE} Trans.
  Inf. Theory}, vol.~52, no.~2, pp. 361--374, 2006.

\bibitem{6239604}
D.~W.~H. Cai, T.~Q.~S. Quek, C.~W. Tan, and S.~H. Low, ``Max-min {SINR}
  coordinated multipoint downlink transmission—duality and algorithms,''
  \emph{{IEEE} Trans. Signal Process.}, vol.~60, no.~10, pp. 5384--5395, 2012.

\bibitem{4489240}
M.~Codreanu, A.~Tolli, M.~Juntti, and M.~Latva-Aho, ``Uplink-downlink {SINR}
  duality via {L}agrange duality,'' in \emph{Proc. IEEE Wireless Commun. Netw.
  Conf. (WCNC)}, 2008, pp. 1160--1165.

\bibitem{7801160}
H.~Sifaou, A.~Kammoun, L.~Sanguinetti, M.~Debbah, and M.-S. Alouini,
  ``Max–min {SINR} in large-scale single-cell {MU-MIMO}: Asymptotic analysis
  and low-complexity transceivers,'' \emph{{IEEE} Trans. Signal Process.},
  vol.~65, no.~7, pp. 1841--1854, 2017.

\bibitem{nocedal2006numerical}
J.~Nocedal and S.~J. Wright, \emph{Numerical Optimization}.\hskip 1em plus
  0.5em minus 0.4em\relax Springer, 2006.

\bibitem{liu2025ris}
Z.~Liu, W.~Yu, and S.~V. Hum, ``{RIS} and beamformer optimization using hybrid
  full-wave analysis in multiuser {MIMO} networks,'' in \emph{Proc. IEEE Int.
  Symp. Antennas Propag. (AP-S)}, 2025, pp. 788--791.

\bibitem{4153898}
F.~Boccardi and H.~Huang, ``Zero-forcing precoding for the {MIMO} broadcast
  channel under per-antenna power constraints,'' in \emph{Proc. IEEE Workshop
  Signal Process. Adv. Wireless Commun. (SPAWC)}, 2006, pp. 1--5.

\bibitem{9684244}
H.~Yu, H.~D. Tuan, E.~Dutkiewicz, H.~V. Poor, and L.~Hanzo, ``{RIS}-aided
  zero-forcing and regularized zero-forcing beamforming in integrated
  information and energy delivery,'' \emph{{IEEE} Trans. Wireless Commun.},
  vol.~21, no.~7, pp. 5500--5513, 2022.

\end{thebibliography}

\end{document}